\documentclass[12pt]{article}
\usepackage{amsmath,amsfonts,epsfig,amssymb}

\newcommand{\dx}{\mbox{d}x}
\newcommand{\dxi}{\mbox{d}\xi}
\newcommand{\dq}{\mbox{d}q}

\newcommand{\dy}{\mbox{d}y}

\newcommand{\dt}{\mbox{d}t}

\newcommand{\picturesAB}[3]{
\centerline{\raise 2mm \hbox{\raise #3 \hbox{(a)}}
\hspace*{-.3in}
\psfig{file=#1,height=#3}
\hspace*{.02in}
\raise 2mm \hbox{\raise #3 \hbox{(b)}}
\hspace*{-.3in}
\psfig{file=#2,height=#3}
}}
\newcommand{\picturesABC}[4]{
\centerline{\raise #4 \hbox{(a)}
\psfig{file=#1,height=#4}
\hspace*{.2in}
\raise #4 \hbox{(b)}
\psfig{file=#2,height=#4}
\hspace*{.2in}
\raise #4\hbox{(c)}
\psfig{file=#3,height=#4}
}}
\numberwithin{equation}{section}

\begin{document}

\title{Gene regulatory networks: a coarse-grained, equation-free 
       approach to multiscale computation} 

\author{
Radek Erban\thanks{University of Oxford, Mathematical Institute, 
24-29 St. Giles', Oxford, OX1 3LB, United Kingdom, 
{\it e-mail:  erban@maths.ox.ac.uk}; corresponding author.}
 \and  \qquad 
Ioannis G. Kevrekidis\thanks{Princeton University,
Department Of Chemical Engineering, PACM and Mathematics,
Engineering Quadrangle,
Olden Street, Princeton, NJ 08544, USA;
{\it e-mail: kevrekidis@princeton.edu}.}
\qquad  \qquad \qquad \quad
\and \qquad  \qquad \quad
David Adalsteinsson\thanks{University of North Carolina, 
Applied Mathematics Program, Department of Mathematics, 
Chapel Hill, NC 27599, USA, {\it e-mail: david@amath.unc.edu}.}
\and \; \qquad Timothy C. Elston\thanks{University of North Carolina,
Department of Pharmacology,  Chapel Hill, NC 27599, 
USA, {\it e-mail: telston@med.unc.edu}.} \quad
}

\date{\today}

\maketitle

{ 
\small \par \noindent 
{\it Abstract:} 
We present computer-assisted methods for analyzing stochastic
models of gene regulatory networks.  
The main idea that underlies this equation-free analysis 
is the design and execution of appropriately-initialized short bursts of
stochastic simulations; the results of these are processed to estimate
coarse-grained quantities of interest, 
such as mesoscopic transport coefficients.
In particular, using a simple model of a genetic toggle switch, 
we illustrate the computation of an effective free energy $\Phi$ 
and of a state-dependent effective diffusion coefficient
$D$ that characterize an unavailable effective Fokker-Planck equation.
Additionally we illustrate the linking of equation-free techniques
with continuation methods for performing a form of
stochastic ``bifurcation analysis"; estimation of mean switching times 
in the case of a bistable switch is also implemented in this equation-free
context.
The accuracy of our methods is tested by direct comparison 
with long-time stochastic simulations. 
This type of equation-free analysis appears to be a promising
approach to computing features of the long-time, coarse-grained behavior of 
certain classes of complex stochastic models of gene regulatory networks,
circumventing the need for long Monte Carlo simulations. 
}

\section{Introduction}

Various ways to model gene-regulatory networks exist,
ranging from logical (boolean), to sto\-chas\-tic (Monte Carlo methods)
or deterministic (ordinary differential equations) models
(for recent reviews 
see \cite{Schlitt:2005:MGN,Hasty:2001:CSG,Kaern:2005:SGE}). 
Each modeling approach has its advantages and
disadvantages.
One advantage of stochastic modeling is that it
takes into account fluctuations due to the inherently random
nature of biochemical reactions.
This intrinsic noise gives rise to significant effects 
when either the molecular abundances 
of protein or mRNA molecules are
small or the kinetics of the
transitions between the chemical states of the promoter are
slow  \cite{Kepler:2001:STR,Kaern:2005:SGE}.

The established approach for stochastic modeling of spatially
homogeneous chemical systems was introduced by
Gillespie \cite{Gillespie:1977:ESS}. 
The Gillespie Stochastic Simulation Algorithm (SSA) is based on
repeatedly answering two questions: when does the next
chemical reaction occur and what kind of reaction is it?
Gillespie \cite{Gillespie:1977:ESS} derived a simple way
to answer these two questions that reduces the problem
to a continuous-time discrete space Markov process.

The SSA generates exact sample paths of the stochastic process and,
for sufficiently large networks, it is computationally more 
efficient than solving the chemical master equation.  
However, the large size of naturally occurring gene regulatory
networks makes even the SSA computationally intensive
and practically impossible to use for computing the long-time behavior of the network.
Consequently, an important restriction of stochastic computations
for many networks of interest is that we can efficiently run 
stochastic Gillespie-based simulators for {\it short times} only. 
It is therefore natural to look for computational methods that 
use only short time simulations (and as few of these as necessary)
to compute the required information for the system. 
Such a computer-assisted approach is presented in this paper.

Model reduction often provides a natural path to efficient
simulation of a complicated model.
As in other branches of physical modeling, separation of 
time scales can lead to successful model reduction in gene 
regulatory network modeling.
Separation of time scales is frequently present in this context
because synthesis and degradation of new proteins and transcripts usually
occurs on a slower time scale than processes that change the chemical
state of
proteins (e.g., multimerization, protein/DNA interactions, phosphorylation).
Theoretical methods for stochastic model
reduction that take advantage of separation of time scales are being
developed (e.g. \cite{Kepler:2001:STR,Cao:2005:SCS,Haseltine:2002:ASC,
Rao:2003:SCK}).
Analytical reduction techniques assume that fast 
variables are in quasi-steady state with respect to the remaining 
slow variables. 
If the quasi-steady state distributions conditioned on the
slow variables can be determined, then they can be used to eliminate the
fast variables.

Our approach is also based on (and takes advantage of) the 
separation of time scales and the approximation (computationally,
on the fly) of quasi-steady marginal distributions (conditioned
on the slow variables).
The main feature of our approach, as will become apparent through
its description and illustration, is that we do not ``first reduce
and then simulate the reduced model"; our methods come in the form
of wrappers around a black box dynamic simulator, and could equally
well be applied to the most detailed stochastic version of the network model 
or to its best explicit reduction already available.
In our approach, results about the long-term dynamic behavior of
a stochastic simulator do not come from long-term simulation; they
come from the design, execution and processing of the results of
``intelligently designed" short bursts of direct dynamic simulation.

We believe it is useful to draw here an analogy with the study of
nonlinear dynamics in systems of ODEs.
Long-term information in the form
of detailed bifurcation diagrams 
can be obtained from long dynamic integration; yet the same
information is much more 
systematically and economically obtained through {\it different algorithms}
using the same model: bifurcation, stability and continuation methods.
It is this alternative to direct, long-term stochastic
simulation (whether with the full detailed network model or with any
good analytical reduction of it) that our approach makes available
to the modeler.
Ours is a ``design of computational experiments" approach; it is
guided by model reduction, but a reduced model is never explicitly
obtained.

The remainder of the paper is organized as follows.
In Section \ref{model}, we introduce the genetic toggle switch
as a simple model to illustrate our methods,
and we specify the main questions that one would like to answer with these techniques.
In Section \ref{secmathfram}, we present
the general mathematical framework and main ideas
of  {\it equation-free}  analysis 
\cite{Kevrekidis:2003:EFM,Erban:2004:CAE,Gear:2002:CIB,Siettos:2003:CBD,Haataja:2004:AHD}.
In Section \ref{secstosist}, 
we present an analysis of a deterministic model of  the genetic toggle
switch to provide insight into this system.
We also introduce several stochastic  models of increasing complexity that are used 
to illustrate equation-free analysis.
In Section \ref{secnumer}, we 
compute the effective free energies and the associated stationary
distributions  for the stochastic models described in Section \ref{secstosist}.
Equation-free bifurcation analysis is then presented, and, in bistable
cases, the mean first passage times for the system to switch 
between apparent stable fixed points are computed.
We end with a discussion of the equation-free approach, its strengths,
weaknesses, relations to other current methods for the acceleration
of SSA-type simulations (e.g. 
\cite{Kepler:2001:STR,Cao:2005:SCS,Haseltine:2002:ASC,
Rao:2003:SCK,Adalsteinsson:2004:BNS}) and its possible 
extensions in Section \ref{secdiscussion}. 
In particular, we will discuss the applicability of our
methods to more complicated gene-regulatory networks.

\section{Model Description}

\label{model}

Our illustrative example is a two gene network in 
which each protein represses the transcription of the other gene 
(mutual repression).
This type of system has been  engineered in {\it E. coli} and is often referred to 
as a genetic toggle switch \cite{Gardner:2000:CGT,Hasty:2002:EGC}.
The advantage of this simple system is that it allows us to
test the accuracy of equation-free methods by direct 
comparisons with results from long-time
stochastic simulations. 
In Section \ref{secdiscussion}, we
discuss the applicability of our methods to more complex
problems where long direct stochastic simulation is impossible
and the accuracy must be checked by on line {\it a posteriori}
error estimates.

\begin{figure}
\centerline{
\psfig{file=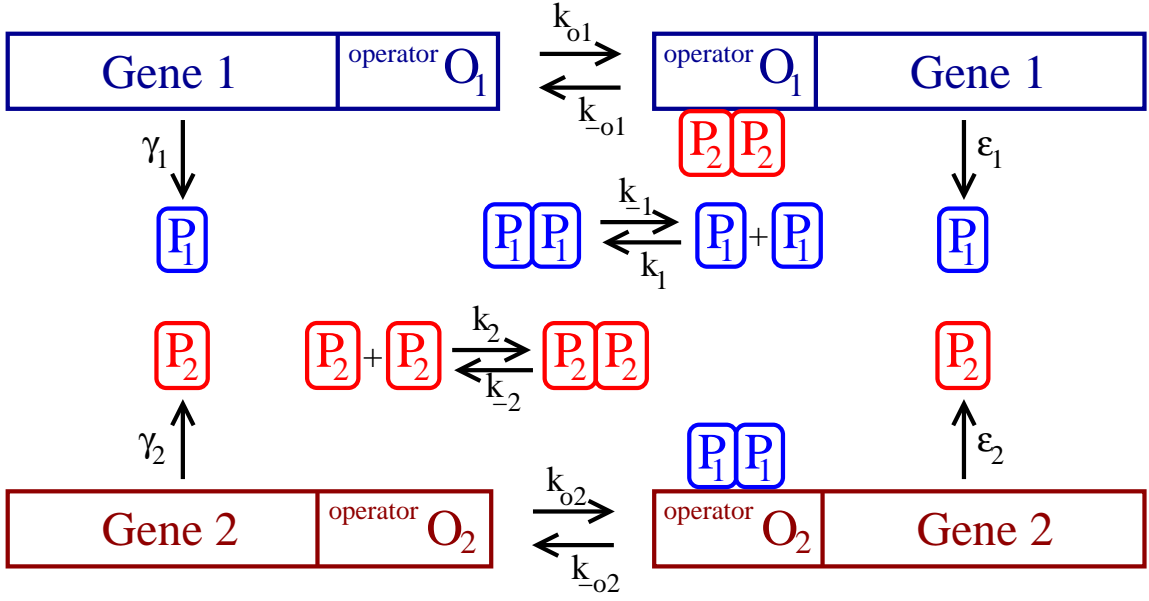,width=6in}
}
\caption{{\it A schematic diagram of the genetic toggle switch.}}
\label{figmutrep}
\end{figure}

A  simple version of the genetic toggle switch is schematically
drawn in Figure \ref{figmutrep}. 
The system contains two proteins $P_1$ and $P_2$. 
The production of $P_1$ ($P_2$) depends on the
chemical state of the upstream operator $O_1$ ($O_2$).
If $O_1$ is empty then $P_1$ is produced at the rate
$\gamma_1$ and if $O_1$ is occupied by a dimer of $P_2$, then protein $P_1$
is produced at a rate $\epsilon_1 < \gamma_1.$
Similarly, if $O_2$ is empty then $P_2$ is produced at the rate
$\gamma_2$ and if $O_2$ is occupied by a dimer of $P_1$, then protein $P_2$
is produced at a rate $\epsilon_2 < \gamma_2.$
Note that for simplicity, transcription and translation
are described by a single rate constant. 
The biochemical reactions 
and rate constants that correspond
to the processes shown in Figure \ref{figmutrep}  are
\begin{equation}
{\mbox{ \raise 0.851 mm \hbox{$\emptyset$}}}
\quad \;
\mathop{\stackrel{\displaystyle\longrightarrow}\longleftarrow}^{\gamma_1 O_1 + \varepsilon_1 \overline{P_2P_2O_1}}_{\delta_1}
\quad
{\mbox{ \raise 0.851 mm \hbox{$P_1$}}}
\label{prodP1}
\end{equation}
\begin{equation}
{\mbox{ \raise 0.851 mm \hbox{$\emptyset$}}}
\quad \;
\mathop{\stackrel{\displaystyle\longrightarrow}\longleftarrow}^{\gamma_2 O_2 + \varepsilon_2 \overline{P_1P_1O_2}}_{\delta_2}
\quad
{\mbox{ \raise 0.851 mm \hbox{$P_2$}}}
\label{prodP2}
\end{equation}
\begin{equation}
{\mbox{ \raise 0.851 mm \hbox{$P_1 + P_1$}}}
\quad \;
\mathop{\stackrel{\displaystyle\longrightarrow}\longleftarrow}^{k_{1}}_{k_{-1}}
\quad
{\mbox{ \raise 0.851 mm \hbox{$\overline{P_1P_1}$}}}
\label{prodP1P1}
\end{equation}
\begin{equation}
{\mbox{ \raise 0.851 mm \hbox{$P_2 + P_2$}}}
\quad \;
\mathop{\stackrel{\displaystyle\longrightarrow}\longleftarrow}^{k_{2}}_{k_{-2}}
\quad
{\mbox{ \raise 0.851 mm \hbox{$\overline{P_2P_2}$}}}
\label{prodP2P2}
\end{equation}
\begin{equation}
{\mbox{ \raise 0.851 mm \hbox{$\overline{P_2P_2}+O_1$}}}
\quad \;
\mathop{\stackrel{\displaystyle\longrightarrow}\longleftarrow}^{k_{o1}}_{k_{-o1}}
\quad
{\mbox{ \raise 0.851 mm \hbox{$\overline{P_2P_2O_1}$}}}
\label{onoffO1}
\end{equation}
\begin{equation}
{\mbox{ \raise 0.851 mm \hbox{$\overline{P_1P_1}+O_2$}}}
\quad \;
\mathop{\stackrel{\displaystyle\longrightarrow}{\longleftarrow}}^{k_{o2}}_{k_{-o2}}
\quad
{\mbox{ \raise 0.851 mm \hbox{$\overline{P_1P_1O_2}$}}}
\label{onoffO2}
\end{equation}
where the overbars denote complexes. 
Equation (\ref{prodP1}) describes production and degradation
of protein $P_1$, equation (\ref{prodP2}) describes
production and degradation of protein $P_2$,
equations (\ref{prodP1P1}) and (\ref{prodP2P2})
are dimerization reactions and equations (\ref{onoffO1}) and
(\ref{onoffO2}) represent the binding and dissociation of the dimer and DNA.

Single cell fluorescence measurements can be used
to measure intercellular variability in protein expression levels.
Therefore it is important  to have efficient methods for computing the
steady-state distribution of protein abundances from stochastic 
models similar to the one defined by (\ref{prodP1}) -- (\ref{onoffO2}). 
For
moderately complex systems using long-time Monte Carlo simulations quickly becomes
computationally prohibitive. 
We will illustrate how equation-free analysis can  
overcome this difficulty by accelerating the exploration of certain features of
the long-term dynamics of the stochastic simulation.
For certain values of the model parameters,
the genetic toggle switch is  bistable. 
If the system is described
in terms of ordinary differential equations (ODEs) for the protein concentrations, 
then standard bifurcation analysis 
(numerical continuation methods) can
be applied to determine the regions of parameter space in which
bistability occurs. 
Using the model described by (\ref{prodP1}) -- (\ref{onoffO2})
as an example,
we show how to extend these techniques
to stochastic models. 
An important quantity that characterizes the dynamics of bistable
stochastic systems is the average
time for spontaneous transitions between stable steady states to occur.
We will illustrate how this mean first passage time can be computed by using only
short-time simulations. 

\section{Equation-Free Analysis: Mathematical Framework}

\label{secmathfram}

Let us suppose that we have a well-stirred mixture of chemically reacting species;
our main assumption is that the evolution of the system
can be described in terms of a single,
slowly evolving random variable $Q$ (the approach carries through for the
case of a small number of slow variables, but in this paper we will focus on
the single slow variable case).
$Q$ might be the  concentration of
one of the chemical species
or some function of the concentrations.
Let $\bf R$ denote a vector of the other (fast, ``slaved" system
variables). 
Our assumption implies that the evolution of the system can be approximately described by
the time-dependent probability density function $f(q,t)$ for the slow variable $Q$
that evolves according to following 
effective Fokker-Planck equation \cite{Risken:1989:FPE}:
\begin{equation}
\frac{\partial f}{\partial t}
=
\frac{\partial}{\partial q}
\left(
\frac{\partial}{\partial q}
[D(q) f(q,t)]
-
V(q) f(q,t)
\right).
\label{FPE}
\end{equation}
If the effective drift $V(q)$ and diffusion coefficient $D(q)$ could be 
explicitly written down as function of $q$, 
then (\ref{FPE}) can be used to compute interesting properties
of the system (e.~g., the steady state distribution). 
Note that
in addition to the assumption of a single slow variable, the validity
of equation (\ref{FPE}) requires sufficiently large molecular abundances
and sufficiently fast chemical kinetics for transitions in the chemical
state of the operator \cite{Gardner:2000:CGT,Kepler:2001:STR}.
Assuming that (\ref{FPE}) provides a good approximation, we
make use of the following formulas for the drift and
diffusion coefficient 
\cite{Haataja:2004:AHD,Hummer:2003:CMD,Kopelevich:2005:CGK,Sriraman:2005:CND}
\begin{eqnarray}
V(q)
&=&
\lim_{\Delta t \to 0}
\frac{ < Q(t+\Delta t) - q \, | \, Q(t)=q>}{\Delta t} \label{avgvel} \\
D(q)
&=&
\frac{1}{2} \,
\lim_{\Delta t \to 0}
\frac{ < [Q(t+\Delta t) - q]^2 \, | \, Q(t)=q>}{\Delta t}.
\label{effdiff}
\end{eqnarray}
As described below, estimates of these two quantities can be found by 
using short-time bursts of  appropriately initialized stochastic simulations. 
The steady solution of (\ref{FPE}) is proportional
to $\exp[-\beta \Phi(q)]$, where the effective
free energy $\Phi (q)$ is defined as
\begin{equation}
\frac{\Phi(q)}{k_BT} \equiv \beta \Phi(q) =
- \int_0^q \frac{V(q^\prime)}{D(q^\prime)} \dq^\prime
+ \ln D(q) + \mbox{constant}.
\label{potentialPhi}
\end{equation}
Consequently, computing the effective free energy and the steady
state probability distribution also can be accomplished without the need for 
long-time stochastic simulations. 

\noindent
A procedure for computationally estimating
$V(q)$ and $D(q)$ is as follows:

\leftskip 1cm
\bigskip

\noindent
{\bf (A)} Given $Q=q$, approximate the conditional density $P({\bf r}|Q=q)$ for the fast variables $\bf R$. 
Details of this preparatory step are given below.

\noindent
{\bf (B)} Use $P({\bf r}|Q=q)$ from the step (A) to determine
appropriate initial conditions for the short simulations and run multiple realizations
for time $\Delta t$. Use the results of these simulations and
the definitions (\ref{avgvel}) and (\ref{effdiff}) to estimate
the average velocity $V(q)$ and effective diffusion coefficient $D(q)$. 

\noindent
{\bf (C)} Repeat steps (A) and (B) for sufficiently many values of $Q$ and then compute
$\Phi(q)$ using formula (\ref{potentialPhi}) and numerical quadrature.

\leftskip 0cm
\bigskip

\noindent
A very important feature of this algorithm is that it is trivially parallelizable 
(different realizations of short simulations starting at ``the same $q$" as
well a simulation realizations starting at different $q$ values
can be run independently, on multiple processors).

In order to use the algorithm (A) -- (C), we have to specify 
how the step (A) is performed. 
There are several computational options to approximate the 
conditional density $P({\bf r}|Q=q).$ 
The simplest approximation is to estimate (through numerical experiments) 
the conditional mean $<\!\!{\bf R}|{Q=q}\!\!>$  and approximate $P({\bf r}|Q=q)$ 
as a Dirac delta function $\delta({\bf r }- <\!\!{\bf R}|Q=q\!\!>)$.
Then the step (A) reads as follows:

\leftskip 1cm
\bigskip

\noindent
{\bf (A1)}  Given $Q=q$, pick an initial guess for the conditional mean 
of ${\bf R}$. 
Denote the initial guess as $<\!\!{\bf R}(0)\!\!>$. Run multiple 
realizations for a short time $\delta t$ and compute 
$<\!\!{\bf R}(\delta t)\!\!>.$  
This procedure defines the mapping 
$<\!\!{\bf R}(0)\!\!> \to <\!\!{\bf R}(\delta t)\!\!>.$ 
Find the steady state 
of this mapping using standard numerical methods. 
The steady state is the required
conditional average $<\!\!{\bf R}|{Q=q}\!\!>.$ 
Initialize ${\bf R}(0)$ as $<\!\!{\bf R}|{Q=q}\!\!>$ in all realizations
in part (B) of the algorithm.

\leftskip 0cm
\bigskip

\noindent
Another option is to approximate $P({\bf r}|Q=q)$  as a distribution characterized
by a few parameters, e.g. as a Gaussian distribution with
mean $\boldsymbol{\mu}$ and variance $\boldsymbol{\sigma}.$ 
This can be done as follows:

\leftskip 1cm
\bigskip

\noindent
{\bf (A2)} Given $Q=q$, pick initial guesses for the mean
$\boldsymbol{\mu}(0)$ and variance $\boldsymbol{\sigma}(0)$ of the conditional 
distribution function $P({\bf r}|Q=q)$. 
Use this distribution 
to generate many realizations of ${\bf R}(0)$. Using these realizations 
as initial conditions, run stochastic simulations for a short time $\delta t$ 
and compute ${\bf R} (\delta t).$ 
Computing mean
and variance of ${\bf R} (\delta t),$ we obtain the mapping
$[\boldsymbol{\mu}(0),\boldsymbol{\sigma}(0)] \to [\boldsymbol{\mu}(\delta t),\boldsymbol{\sigma}(\delta t)].$ 
Next use
standard numerical methods to find the steady state 
$[\boldsymbol{\mu},\boldsymbol{\sigma}]$ of this mapping and 
approximate $P({\bf r}|Q=q)$ 
as a Gaussian distribution with mean $\boldsymbol{\mu}$ and 
variance $\boldsymbol{\sigma}.$

\leftskip 0cm
\bigskip

\noindent
The conditional density $P({\bf r}|Q=q)$ can be also approximated
by other basis functions. 
It is straightforward to generalize (A1) 
or (A2) to such a case. 
The better the approximation of $P({\bf r}|Q=q)$ we have,
the shorter the time step, $\delta t$, required in the step (B)  to achieve the
same accuracy. 
So, a better approximation of $P({\bf r}|Q=q)$  in step (A)
decreases the computational intensity of step (B). On the other hand, 
step (A) is  more computationally intensive if we want to obtain
a better approximation of $P({\bf r}|Q=q)$. 
One possibility for generating
a better approximation of   $P({\bf r}|Q=q)$ is to use a ``run-and-reset" procedure 
as was done in \cite{Erban:2004:CAE}. 
This is accomplished as follows.

\leftskip 1cm
\bigskip

\noindent
{\bf (A3)} Given $Q=q$, initialize the other variables ${\bf R} \equiv {\bf R}(0)$ of the
system. Run stochastic simulations for the short time $\delta t$. 
Then reset the value
of $Q(\delta t)$ to its original value $q$ keeping ${\bf R}$ unchanged.
Repeat this procedure for many time steps and compute the 
conditional density $P({\bf r}|Q=q)$ as a histogram of 
the recorded values of ${\bf R}$. 

\leftskip 0cm
\bigskip

\noindent
The approach (A3) attempts to compute the $P({\bf r}|Q=q)$ 
effectively by successive substitution, without resorting to
numerical algorithms
of the Newton-Raphson type for locating fixed points of mappings;
we will return to this latter issue in the Discussion section.
In our illustrative computations in Section \ref{secnumer},
we use step (A) in the form (A1) or (A3) for the simple stochastic models
described below. 
Both give good results for our illustrative example.
Since (A1) works for sufficiently long times $\delta t$, there is no
need to use (A2) or higher order approximations.  
For some stochastic simulations of our model problem, we also use slightly 
modified versions
of the methods (A1) or (A3) as will be described in Section \ref{secnumer}.

\subsection{Bifurcations}

\label{secbifth}

In deterministic problems, we often summarize the parametric dependence
of the long-term dynamics in terms of bifurcation diagrams; for example,
we may plot the steady states of a deterministic set of ODEs as a function
of a distinguished {\it bifurcation parameter}.
Several excellent
continuation methods have been developed, implemented and made
available for this purpose over the years, such as AUTO 
\cite{Doedel:1991:NAC,Doedel:1991:NAC2}.

Here we illustrate how these methods can be extended to stochastic models
\cite{Kevrekidis:2003:EFM,Gear:2002:CIB,Sriraman:2005:CND,Makeev:2002:CSB,Makeev:2002:CBA}.
We assume, as above, that we have a stochastic problem
that can be {\it effectively} described by a single variable $Q.$
Let $\gamma$ be the bifurcation parameter.
The first two steps in the algorithm are as follows:

\bigskip

\leftskip 1cm

\noindent
{\bf ($\mathbb A$)} Given $Q = q$ and the value of the bifurcation 
parameter $\gamma$, compute the conditional density  $P({\bf r}|Q=q)$ 
using step (A) of the previous algorithm.

\noindent
{\bf ($\mathbb B$)} Using $P({\bf r}|Q=q)$ from step ($\mathbb A$) to
      determine the initial conditions,
      run multiple stochastic simulations for a short time $\Delta t$
      and compute the conditional average 
      $<\!\!Q(\Delta t)|Q(0) = q\!\!>$.

\bigskip

\leftskip 0cm

\noindent
Steps $\mathbb A$ and $\mathbb B$ define the mapping $(Q(0),\gamma) \to
 <\!\!Q(\Delta t)\!\!>$.
 We denote this mapping as $F$, i.e. $F(Q,\gamma) = <\!\!Q(\Delta t)\!\!>.$
 Our goal is to track the fixed points of $F$ (i.e. $F(Q,\gamma) = Q$)
 as the bifurcation parameter $\gamma$ is varied. 
To do this,
 we first use a Newton-Raphson algorithm to find two steady states
 $(Q_1,\gamma_1)$ and $(Q_2,\gamma_2)$ which are sufficiently close
 to each other (note that one can estimate the derivative of $F(Q,\gamma)$
 numerically by evaluating  $F(Q,\gamma)$ at different points). 
 Then, in a parameter continuation context,
 we choose a small parameter $\delta$ (which can be modified
 adaptively during the computation) and find the next steady
 state using continuation. 
That is, the next values
 of $Q$ and $\gamma$ are found by solving the following system of equations
 \begin{equation}
  \left\{
  \begin{array}{ccc}
  Q - F(Q,\gamma) & = & 0, \\
  (Q-Q_2)(Q_2-Q_1) + (\gamma - \gamma_2)(\gamma_2 - \gamma_1) - \delta & = & 0.
  \end{array}
  \right.
 \label{numcont}
 \end{equation}
 To find the solution of (\ref{numcont}), we estimate the Jacobian numerically by
 evaluating $F(Q, \gamma)$ at several points and then
 use Newton-Raphson algorithm.
When the number of variables starts becoming large, matrix-free methods of
iterative numerical linear algebra (such as Broyden, or 
Newton-Krylov GMRES \cite{Kelley:2003:SNE})
can be used to solve for the fixed point, as opposed to full numerical Jacobian
estimation.
The fixed points computed this way provide, under certain conditions, good
estimates of the {\it critical points} (minima, saddles) of the effective potential
$\Phi(q)$) as a function of a model parameter $\gamma$; this issue is
discussed extensively in 
\cite{Makeev:2002:CSB,Makeev:2002:CBA,Barkley:2005:NDD,Sriraman:2005:CND}, 
and we will return to it again in the Discussion section.

\subsection{First Passage Time}

\label{mathfbt}

\begin{figure}
\centerline{
\psfig{file=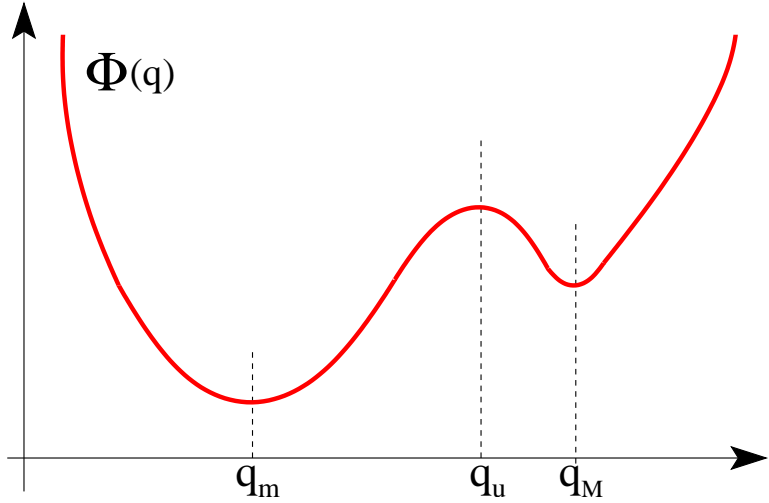,width=4in}
}
\caption{{\it
Potential $\Phi(q)$ of the bistable system.}}
\label{figilpotential}
\end{figure}

Suppose that we have a bistable stochastic system. 
That is, the effective 
free energy $\Phi(q)$ has 
two local minima \cite{Gillespie:1992:MPI} - see Figure \ref{figilpotential}.
Then an important quantity characterizing the long-time system dynamics
is the mean time for spontaneous transitions to occur between
the stable steady states. 
Let $q_m < q_M$ denote the two stable steady
states and let $q_u$ be the unstable state (i.e. local maximum of $\Phi(q)$). 
Then we define the first passage time for transitions from 
$q_m$ to $q_M$ as $2 \tau_{e}$ where $\tau_e$ is the
average time for the system to reach the unstable steady state
$q_u$ for the first time given that it starts at $q_m$.
The factor of two occurs because once the system
reaches the unstable steady state, half the time it returns the original
stable steady state $q_m$ and the other half of the time it transitions to
$q_M$. 

Algorithm (A) -- (C) gives a procedure to estimate the effective 
potential $\Phi(q)$ by running short simulations only.
Once we have the effective potential, we can compute $\tau_e$ 
as follows \cite{Gillespie:1992:MPI}
\begin{equation}
\tau_{e;p} = \int_{q_m}^{q_u} \exp \left[\frac{\Phi(q)}{k_BT} \right]
\int_{-\infty}^q \frac{1}{D(\xi)} \exp \left[- \frac{\Phi(\xi)}{k_BT} \right]
\dxi \dq
\label{formulatauep}
\end{equation}
Equation (\ref{formulatauep}) can be further simplified if the height
of the potential barrier $[\Phi(q_u) - \Phi(q_m)]$
is large compared to the noise strength. 
In this case, the limit
$q$ in the second integral can be replaced by
$q_s$, allowing the two integrals to be evaluated separately
$$
\tau_{e;p} \approx \int_{q_m}^{q_u} \exp \left[\frac{\Phi(q)}{k_BT} \right] \dq
\times
\int_{-\infty}^{q_u} \frac{1}{D(q)} \exp \left[- \frac{\Phi(q)}{k_BT}
\right] \dq.
\label{inteq}
$$
The main contribution of the first integral stems from the
region around $q_s$, and the main contribution from the second
integral stems from the region around $q_m$. 
Consequently,
we expand $\Phi(q)$ according to
$$
\Phi(q) \approx \Phi(q_u) - \frac{1}{2} |\Phi^{\prime\prime} (q_u)|
(q-q_u)^2,
\qquad
\Phi(q) \approx \Phi(q_m) +
\frac{1}{2} \Phi^{\prime\prime} (q_m)
(q-q_m)^2.
$$
for the first and the second integral, respectively
\cite{Risken:1989:FPE}.
When these expansions are used in equation (\ref{inteq}), the following
result is obtained
\begin{equation}
\tau_{e;k} \approx 
\frac{4 \pi k_B T}
{[D(q_u)+D(q_m)]
\sqrt{\Phi^{\prime\prime} (q_m)|\Phi^{\prime\prime} (q_u)|}}
 \exp \left[\frac{\Phi(q_u) -\Phi(q_m) }{k_BT} \right]
\label{tauKramer}
\end{equation}
which is the generalization of Kramers' formula to the case of
a state dependent diffusion coefficient \cite{Risken:1989:FPE,Haataja:2004:AHD}.
Formulas (\ref{formulatauep}) and (\ref{tauKramer}) are both used in 
Section \ref{secexittimes} to estimate $\tau_e.$

\section{Analysis of the Model Problem}

\label{secstosist}

In this section, we study the behavior of the model given by equations
(\ref{prodP1}) -- (\ref{onoffO2}). 
To provide insight into the problem,
we start by analyzing {\it the deterministic system}. 
In Sections
\ref{secmodelA} and \ref{secmodelB}, we introduce
two stochastic models that are simplified versions of the model
defined by (\ref{prodP1}) -- (\ref{onoffO2}). 
We use these models
because of the relative ease in performing long-time stochastic simulations 
with them; 
this allows the results from the equation-free
analysis to be validated by direct comparisons with Monte Carlo simulations.
We will also verify that the equation-free methods can be applied to the full
model. 
As discussed below, for this case the long-time Monte Carlo simulations become
computationally very expensive.

\subsection{The Deterministic Model}
To simplify the deterministic analysis, we make the assumption that equations 
(\ref{prodP1P1}) -- (\ref{onoffO2}) are at quasi-equilibrium and 
derive deterministic rate equations for the protein concentrations.
Let $x_1$ and $x_2$ denote the average monomer 
concentrations of $P_1$ and $P_2$, respectively, and let 
$d_1$ and $d_2$ denote the respective dimer concentrations. 
Also, let  $o_1$ and
$o_2$ denote the probabilities
that the operators $O_1$ and $O_2$ are not occupied. For the
dimerization process
the assumption of quasi-equilibrium implies
\begin{equation}
d_1 = \frac{k_1}{k_{-1}} x_1^2,
\qquad
\mbox{and}
\qquad
d_2 = \frac{k_2}{k_{-2}} x_2^2.
\label{qeq34}
\end{equation}
Similarly, the quasi-equilibrium assumption for the operators implies that
\begin{equation}
o_1 = \frac{k_{-o1}}{k_{-o1} + k_{o1} d_2},
\qquad
\mbox{and}
\qquad
o_2 = \frac{k_{-o2}}{k_{-o2} + k_{o2} d_1}.
\label{qeq56}
\end{equation}
The total concentration of $P_1$ is given by
$y_1 = x_1 + 2 d_1$.
The total concentration $y_1$ evolves according to
the following ordinary differential equation
\begin{equation}
\frac{\dy_1}{\dt}
=
\gamma_1 o_1 + \varepsilon_1 (1 - o_1) - \delta_1 x_1.
\label{eq1a}
\end{equation}
where $\delta$ is the degradation rate of the monomers, and
it has been assumed that dimers are protected from degradation.
Substituting $y_1 = x_1 + 2 d_1
= x_1 + 2 \frac{k_1}{k_{-1}} x_1^2$ into (\ref{eq1a}), we obtain
$$
\left(
1 + 4 \frac{k_1}{k_{-1}} x_1
\right)
\frac{\dx_1}{\dt}
=
\gamma_1 o_1 + \varepsilon_1 (1 - o_1) - \delta_1 x_1
$$
Finally, using (\ref{qeq34}) -- (\ref{qeq56}) produces
\begin{equation}
\frac{\dx_1}{\dt}
=
\frac{1}
{
1 + \kappa_1 x_1
}
\left[
\gamma_1 \frac{1}{1 + \omega_1 x_2^2}
+
\varepsilon_1 \frac{\omega_1 x_2^2}{1 + \omega_1 x_2^2}
- \delta_1 x_1
\right]
\label{eq1b}
\end{equation}
where the parameters $\kappa_1$ and $\omega_1$ are defined as follows
\begin{equation}
\kappa_1 = 4 \frac{k_1}{k_{-1}},
\qquad
\mbox{and}
\qquad
\omega_1 = \frac{k_{o1}}{k_{-o1}} \frac{k_2}{k_{-2}}.
\label{defko1}
\end{equation}
Using similar reasoning an analogous equation for $x_2$ can be derived.

For simplicity, we will present the symmetric case in which the
rate constants for processes involving $P_1$ are identical
to those involving $P_2$. 
That is, we assume 
$\kappa \equiv \kappa_1 =  \kappa_2,$ $\gamma \equiv \gamma_1 = \gamma_2,$
$\omega \equiv \omega_1 = \omega_2,$ and $\delta \equiv \delta_1 = \delta_2$.
Moreover, we assume that the production rate is zero
if an operator is occupied, i.e. $\varepsilon_1 = \varepsilon_2 = 0.$
Making these assumptions, (\ref{eq1b}) simplifies to
\begin{equation}
\frac{\dx_1}{\dt}
=
\frac{1}
{
1 + \kappa x_1
}
\left[
\frac{\gamma}{1 + \omega x_2^2}
- \delta x_1
\right],
\label{eq1c}
\end{equation}
and the equation for $x_2$ is obtained by alternating the subscripts in 
the above equation. 
Hence, the problem has been reduced to a system of two equations
with four parameters.
\begin{figure}
\centerline{
\psfig{file=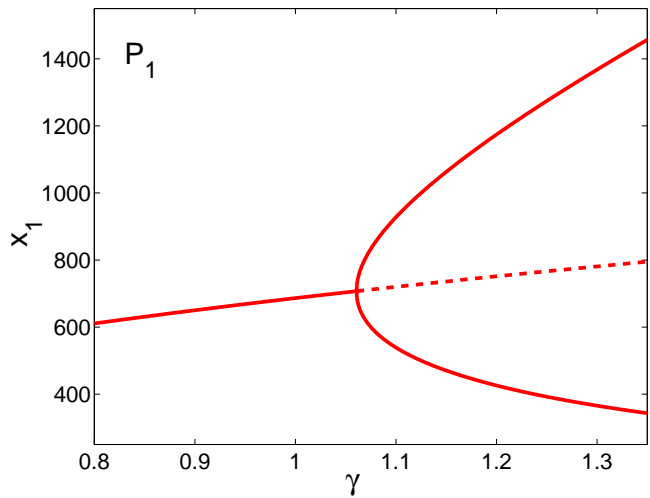,height=2.6in}
}
\caption{{\it The dependence of the steady state values of $x_1$ on $\gamma$.
The solid lines denote stable fixed points and the dashed line corresponds to 
unstable fixed points. 
In this figure and throughout the paper $\delta = 0.00075$
and $\omega = 2 \times 10^{-6}.$ }
}
\label{figP1gammadeter}
\end{figure}
Note that the value of $\kappa$ does not influence the steady-state
behavior of the system.
In this paper, we fix the values of $\delta$ and $\omega$ to be
$0.00075$ and $2 \times 10^{-6}$, respectively.

The steady states values of $x_1$ as a function of $\gamma$ are shown in
Figure \ref{figP1gammadeter}. 
In this figure, solid
lines denote stable steady states and dashed lines denote unstable
steady states.
For $\gamma < 1.06$ there is a single steady state. At $\gamma = 1.06$
a pitchfork bifurcation occurs, and for $\gamma > 1.06,$ there exist three steady
states. 
The steady state with $x_1 = x_2$ is unstable and the other two
steady states are stable. 

Due to separation of time scales, the long-term dynamics of this problem
lie on a lower-dimensional (here one-dimensional) {\it slow manifold}; 
this suggests that one may be able to construct an effective one-dimensional
dynamical system describing the long-term evolution of the model on (near)
this slow manifold.
In constructing such a reduced model, an important question even in the
simple deterministic case is the choice of the right {\it observable} - 
the variable in terms of which the long-term dynamics will be expressed.
An extensive discussion of the choice of such a ``right observable" for
the deterministic case can be found, for example, in \cite{Gear:2004:PSM};
as discussed there, even if we do not know the {\it exact} slow variables,
any set of variables that parametrizes the slow manifold can be practically
used to reduce the system in an equation-free context.
For the stochastic case, a good early illustration and discussion of
manifold parametrization can be found in 
\cite{Siettos:2003:CBD}.
Choosing the right observable is an important issue in the implementation
of equation-free computations, and the subject of intense current research
which we will briefly comment on in Section \ref{secdiscussion}.

\begin{figure}
\centerline{
\psfig{file=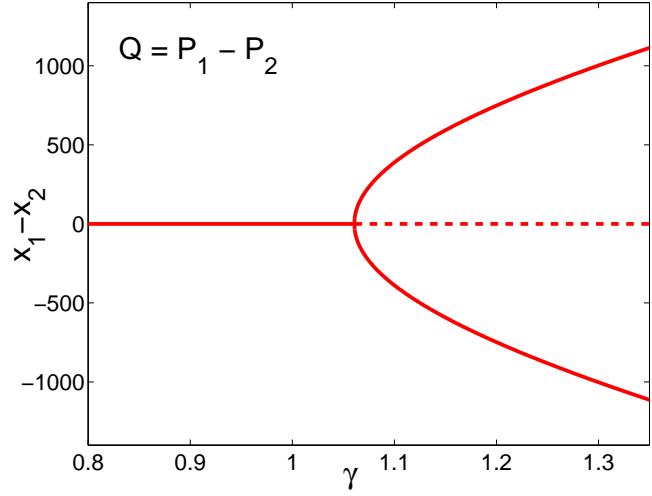,height=2.6in}
}
\caption{{\it
 The dependence of steady state values of the symmetric variable 
 $s = x_1-x_2$ on $\gamma$.}}
\label{figQgammadeter}
\end{figure}

In this paper, and for this example, our equation-free analysis 
assumes that the problem
can be described in terms of a single variable. 
Consequently, it becomes important to select a good observable that
further simplifies the two-dimensional problem to
one dimension.
A tempting (and obvious) choice for the
one-dimensional observable is the molecular abundance of $P_1$ (or $P_2$). 
We demonstrate
below that using $P_1$ in the equation-free analysis produces good
results.
However, we also make use of the symmetric variable defined
as the difference in the protein abundances $Q = P_1 -  P_2$. 
In terms 
of the rate equations the symmetric variable is $s = x_1 - x_2.$
The bifurcation diagram in terms of $s$ is shown in Figure \ref{figQgammadeter}.
The symmetry of the diagram suggests that $Q$ might be a more natural
observable than $P_1$ (which also produces good results, as we will
see below).

\subsection{Stochastic Model I}

\label{secmodelA}

To start our investigations in the equation-free framework, we constructed a 
very simple stochastic model of the system. 
We use this simple model to benchmark equation-free computations, since
the results can be tested against Monte-Carlo simulations easily. 
Results for the full system are also presented below. 
The simple stochastic model 
consists only of reactions for the synthesis and degradation of proteins
$P_1$ and $P_2$, but the following effective rate constants are used
 \begin{equation}
{\mbox{ \raise 1.3 mm \hbox{$\emptyset$}}}
\quad \;
\mathop{\stackrel{\mbox{\Large $\longrightarrow$}}
{\mbox{\Large $\longleftarrow$}}}^{\frac{1}{1 + \kappa P_1} \frac{\gamma}{1 + \omega P_2^2}}_{\frac{\delta}{1 + \kappa P_1}}
\quad
{\mbox{ \raise 1.3 mm \hbox{$P_1$}}}
\label{prodP1s}
\end{equation}
\begin{equation}
{\mbox{ \raise 1.3 mm \hbox{$\emptyset$}}}
\quad \;
\mathop{\stackrel{\mbox{\Large  $\longrightarrow$}}
{\mbox{\Large  $\longleftarrow$}}}^{\frac{1}{1 + \kappa P_2} \frac{\gamma}{1 + \omega P_1^2}}_{\frac{\delta}{1 + \kappa P_2}}
\quad
{\mbox{ \raise 1.3 mm \hbox{$P_2$}}}
\label{prodP2s}
\end{equation}
The above reactions are consistent with the deterministic model, 
but in general do not preserve the noise structure of the full stochastic model. 

\begin{figure}
\centerline{{\large $\gamma = 0.8$}}
\smallskip
\centerline{
\psfig{file=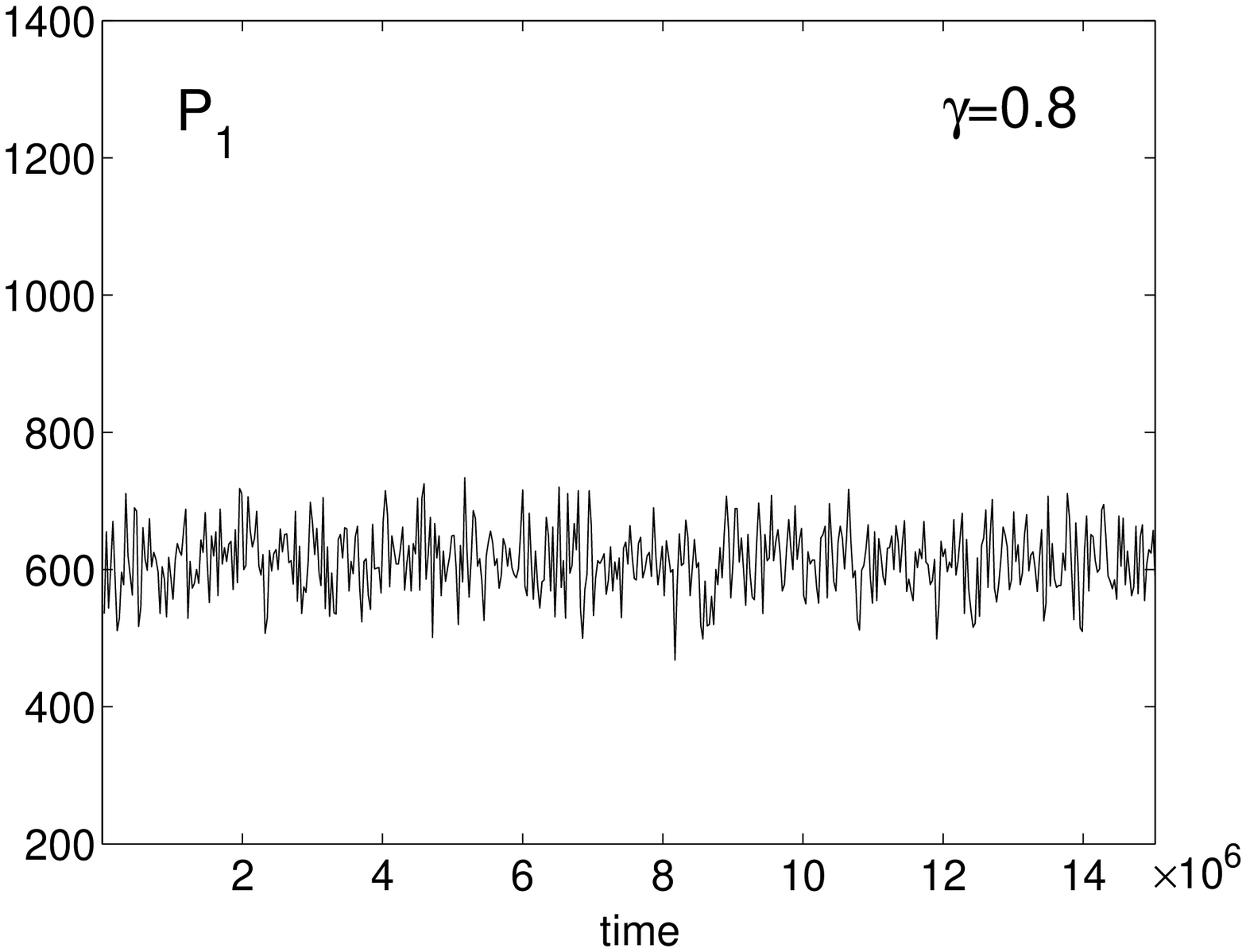,height=2.1in}
\quad
\psfig{file=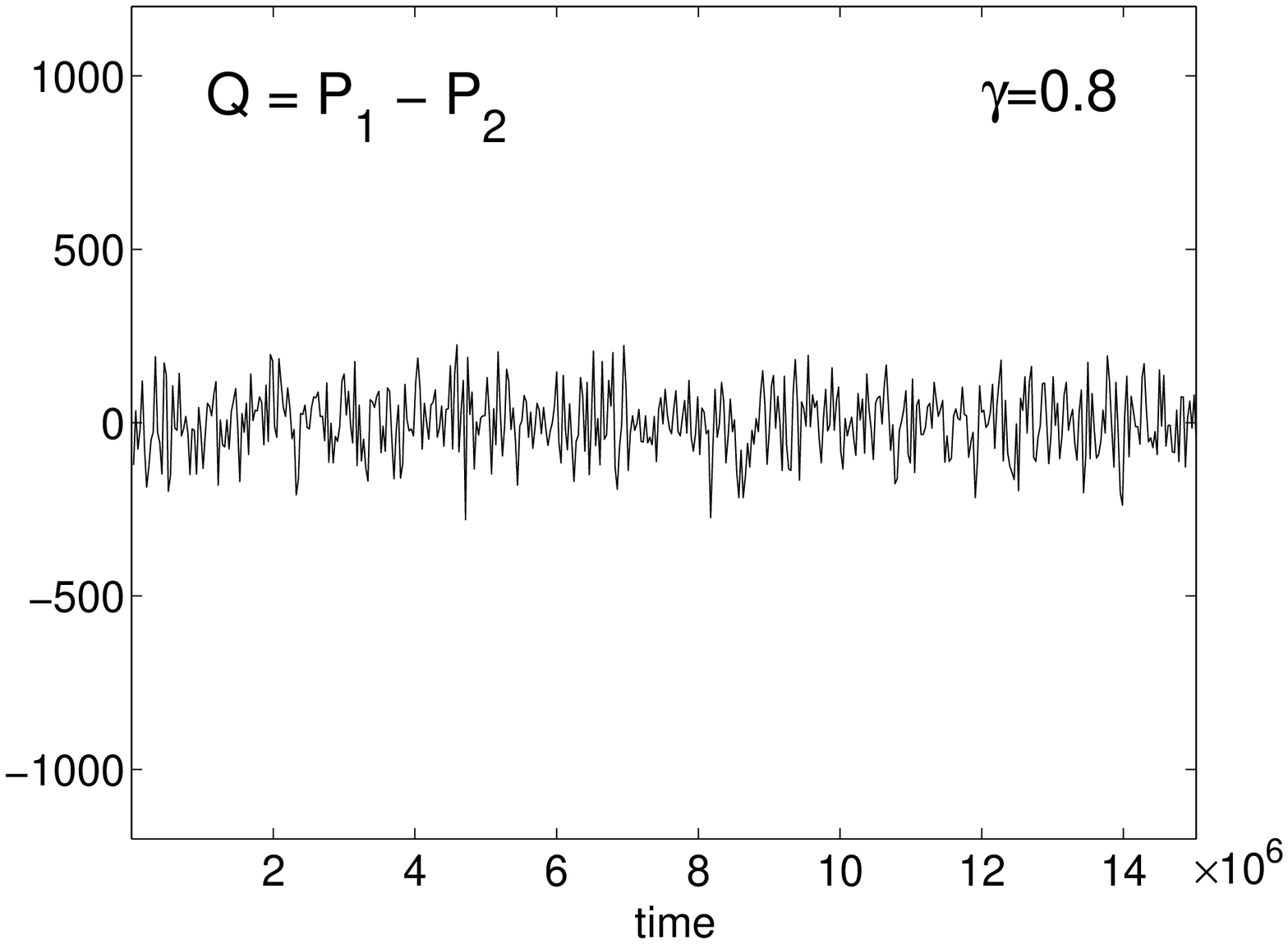,height=2.1in}
}
\bigskip
\centerline{{\large $\gamma = 1.06$}}
\smallskip
\centerline{
\psfig{file=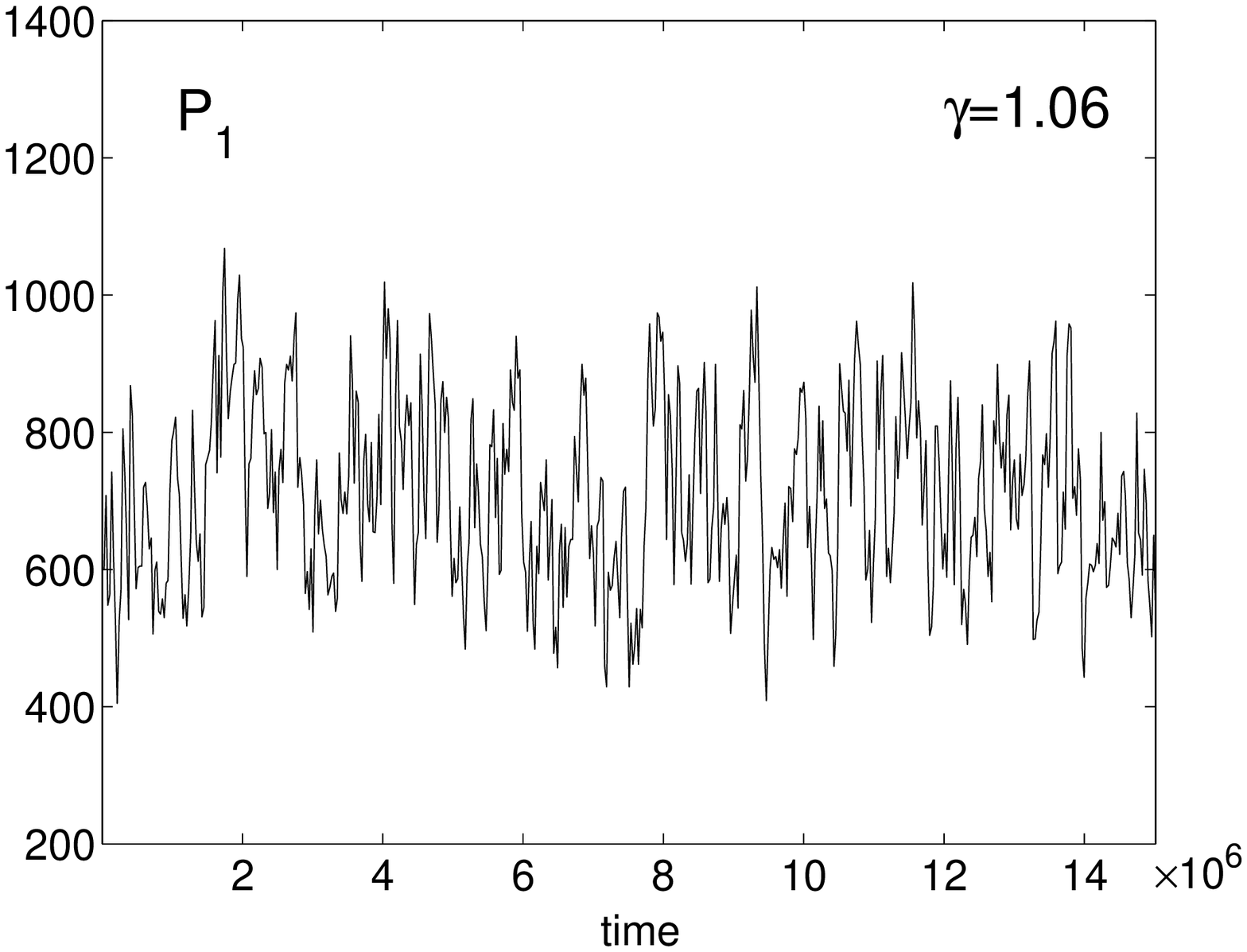,height=2.1in}
\quad
\psfig{file=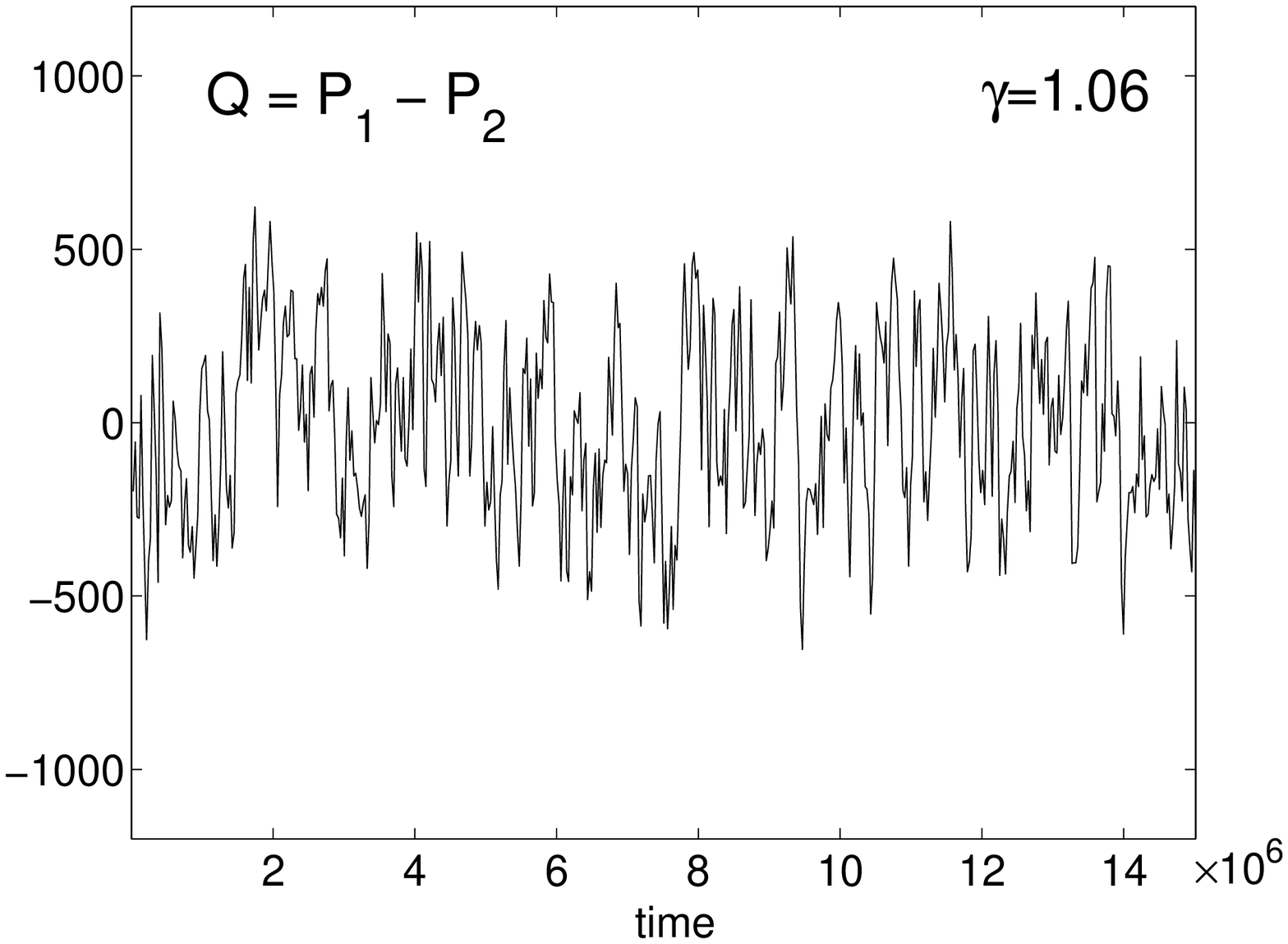,height=2.1in}
}
\bigskip
\centerline{{\large $\gamma = 1.14$}}
\smallskip
\centerline{
\psfig{file=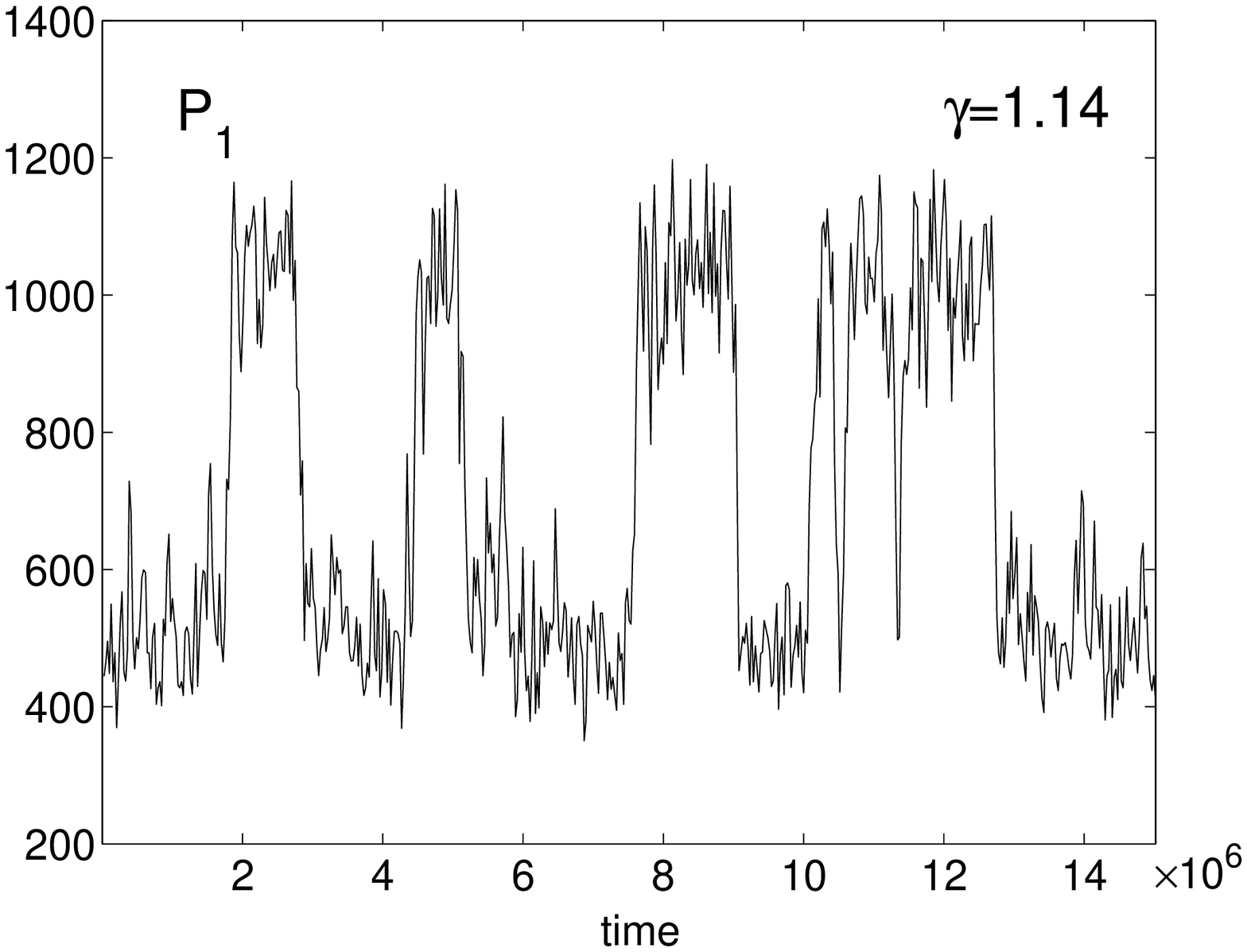,height=2.1in}
\quad
\psfig{file=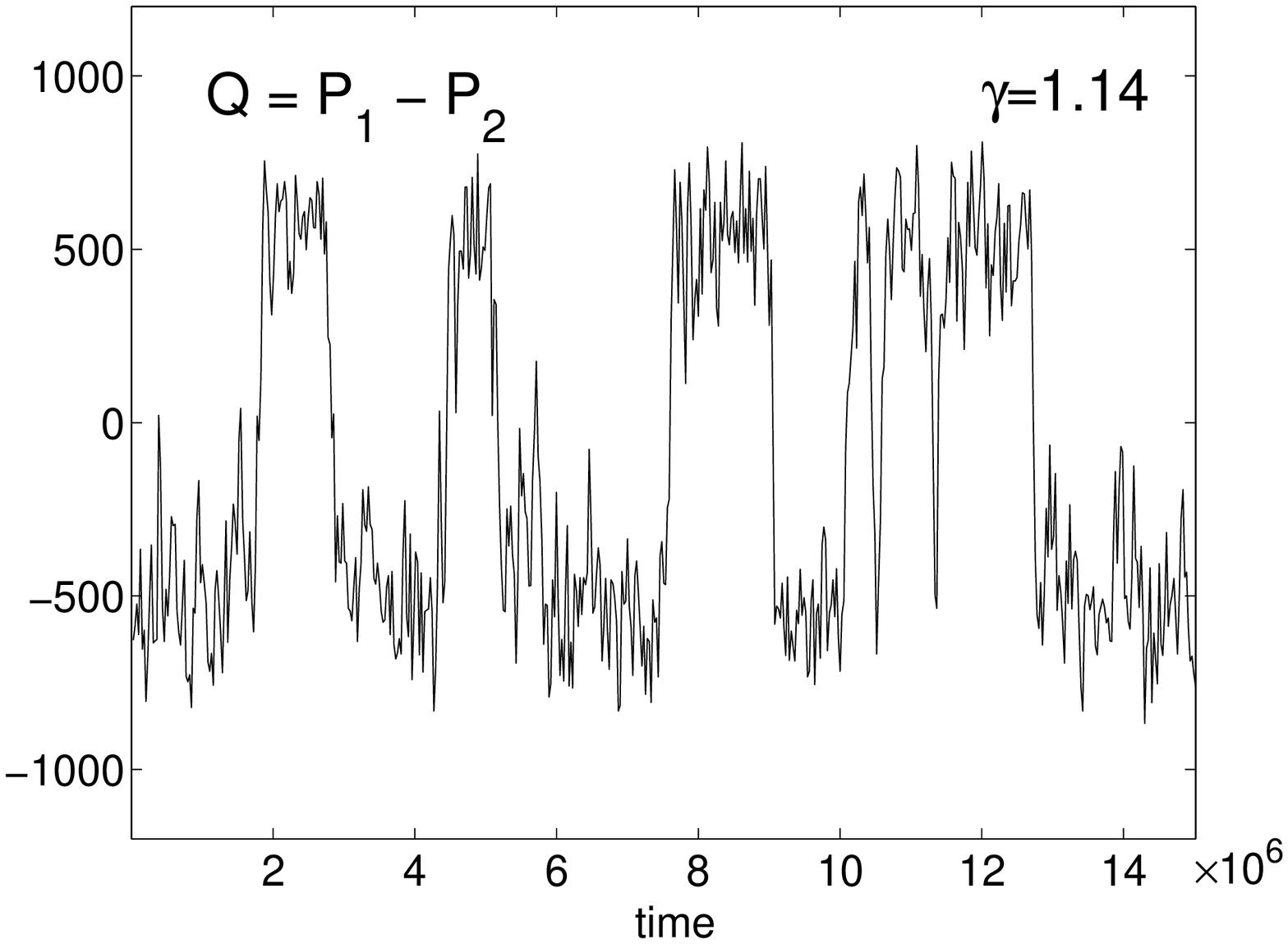,height=2.1in}
}
\caption{{\it
Stochastic Model I. Plots of $P_1$ and $Q=P_1 - P_2$ as a function
of time for different values of $\gamma.$ The parameter values
used to produce these figures are $\delta = 0.00075$,
$\omega = 2 \times 10^{-6}$, and 
$\kappa = 2 \times 10^{-4}$.
}}
\label{figPQ1}
\end{figure}

To simulate the mechanism contained in model (\ref{prodP1s}) -- (\ref{prodP2s}), we
use the standard Gillespie SSA \cite{Gillespie:1977:ESS}.
The results for different values of the parameter $\gamma$ 
are plotted in Figure \ref{figPQ1}.
For each $\gamma$ we plot the time evolution
$P_1$ (left panel) and $Q=P_1-P_2$ (right panel).
We see that for small $\gamma$, the solution fluctuates
around the stable deterministic steady state with relatively small noise
amplitude.
When $\gamma = 1.06$, the noise amplitude has increased substantially,
which is typical of stochastic systems near a ``bifurcation"; the 
word bifurcation is put here in quotes to denote that (in contrast
to the deterministic case) there is no isolated parameter value
marking the onset of bistability - no clear bifurcation point exists 
{\it for the stochastic dynamics}.
Yet one can still claim that a clear bifurcation point exists for
the critical points of the potential $\Phi(q,\gamma)$ in the stochastic
model; furthermore,
depending on the {\it time horizon} of our observation of a stochastic
simulation, one may still appear to see an apparent bifurcation point 
for its averaged {\it statistics} (see the discussion in 
\cite{Haataja:2004:AHD,Barkley:2005:NDD}).
If $\gamma$ is increased further, then $Q=0$ is no longer a stable
steady state and the system clearly shows bistability. 
All plots
are computed for the same time interval $[0,15 \times 10^6].$
For $\gamma=1.25$ the steady states are sufficiently stable so that
no transitions occurred in this time interval (data not shown). 
Therefore, 
for this case, determining
the steady state probability distribution from long-term
Monte-Carlo simulations would be very time 
consuming.

\subsection{Stochastic Model II}

\label{secmodelB}

Stochastic Model I considers only 
two variables $P_1$ and $P_2$. 
Here, we introduce a stochastic
model that also takes into account the biochemical states of the operators, while 
maintaining the assumption that the dimerization
reactions (\ref{prodP1P1}) -- (\ref{prodP2P2}) are  
at equilibrium.
That is, we consider the four variables $P_1$, $P_2$,
$O_1$ and $O_2$.
The model is defined in terms of the following reaction steps:
\begin{equation}
{\mbox{\raise 1.3mm \hbox{$\emptyset$}}}
\quad
\mathop{\stackrel{\mbox{ $\longrightarrow$}}
{\mbox{ $\longleftarrow$}}}^{\frac{\gamma}{1 + \kappa P_1} \, O_1}_{\frac{\delta}{1 + \kappa P_1}}
\quad
{\mbox{ \raise 1.3mm \hbox{$P_1$}}}
\label{prodP1b}
\end{equation}
\begin{equation}
{\mbox{ \raise 1.3mm \hbox{$\emptyset$}}}
\quad
\mathop{\stackrel{\mbox{ $\longrightarrow$}}
{\mbox{ $\longleftarrow$}}}^{\frac{\gamma}{1 + \kappa P_2} \, O_2}_{\frac{\delta}{1 + \kappa P_2}}
\quad
{\mbox{ \raise 1.3mm \hbox{$P_2$}}}
\label{prodP2b}
\end{equation}
\begin{equation}
{\mbox{ \raise 1.3mm \hbox{``$O_1 = 0$"}}}
\quad
\mathop{\stackrel{\mbox{ $\longrightarrow$}}
{\mbox{ $\longleftarrow$}}}^{K}_{K \omega P_2^2}
\quad
{\mbox{\raise 1.3mm \hbox{``$O_1 = 1$"}}}
\label{prodO1b}
\end{equation}
\begin{equation}
{\mbox{ \raise 1.3mm \hbox{``$O_2 = 0$"}}}
\quad
\mathop{\stackrel{\mbox{ $\longrightarrow$}}
{\mbox{ $\longleftarrow$}}}^{K}_{K \omega P_1^2}
\quad
{\mbox{ \raise 1.3mm \hbox{``$O_2 = 1$"}}}
\label{prodO2b}
\end{equation}
and contains an extra parameter, $K \equiv  k_{-o1} = k_{-o2}$. 
Note that ``$O_1 = 0$" means that the operator
$O_1$ has a dimer of $P_2$ bound to it and therefore is 
``off" and ``$O_1 = 1$" means that the operator $O_1$
is empty and therefore ``on". 
The same is true for $O_2.$ 
This implies that the random variables $O_1$
and $O_2$ are binary, whereas the variables $P_1$ and $P_2$
can take on any non-negative integer value. 
Stochastic Model I is recovered from 
Stochastic Model II in the limit $K \to \infty.$ 
We thus expect 
the models to produce similar results for large values of $K.$ 

\begin{figure}
\centerline{{\large $K= 0.1$, $\gamma = 1.14$}}
\smallskip
\centerline{
\psfig{file=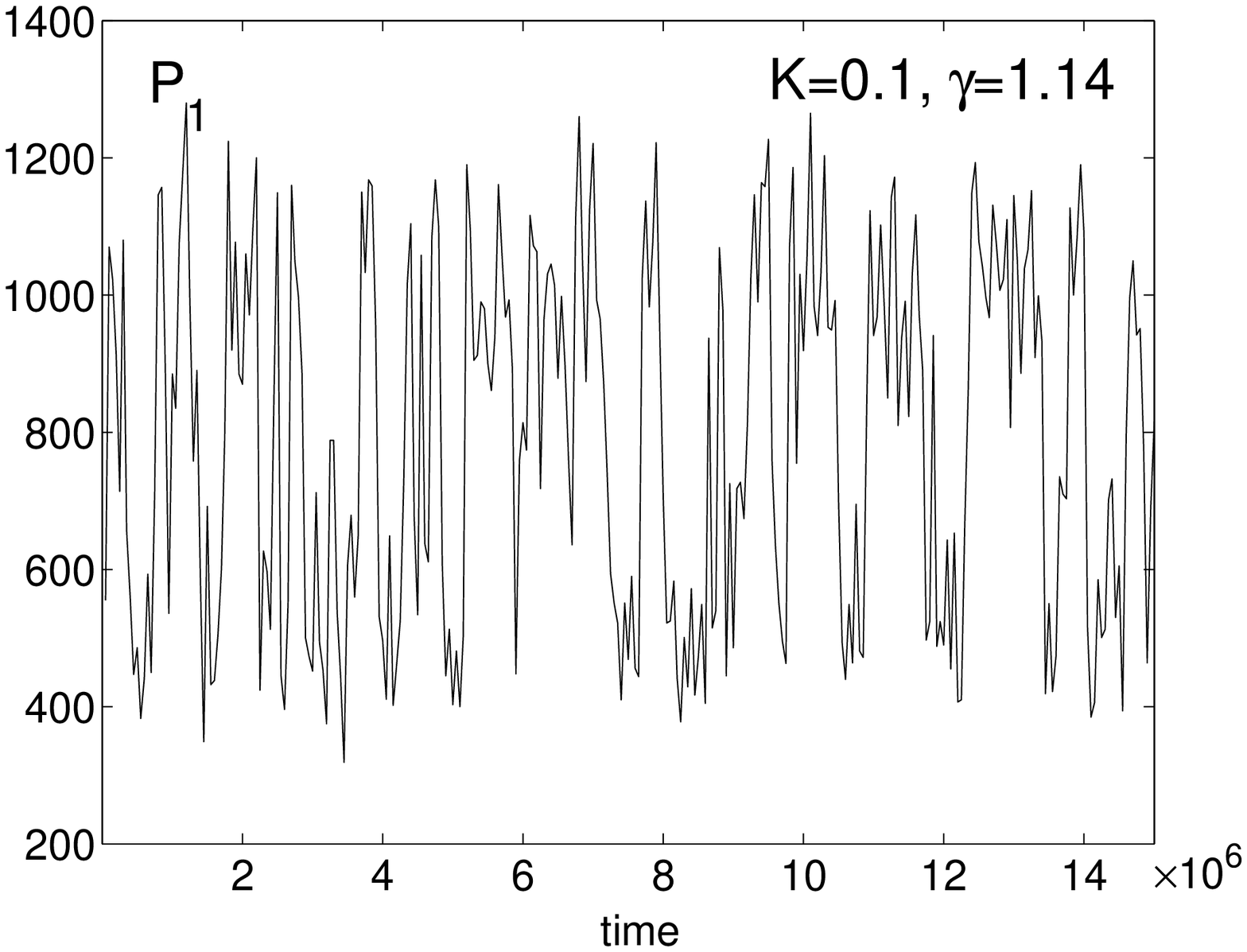,height=2.1in}
\quad
\psfig{file=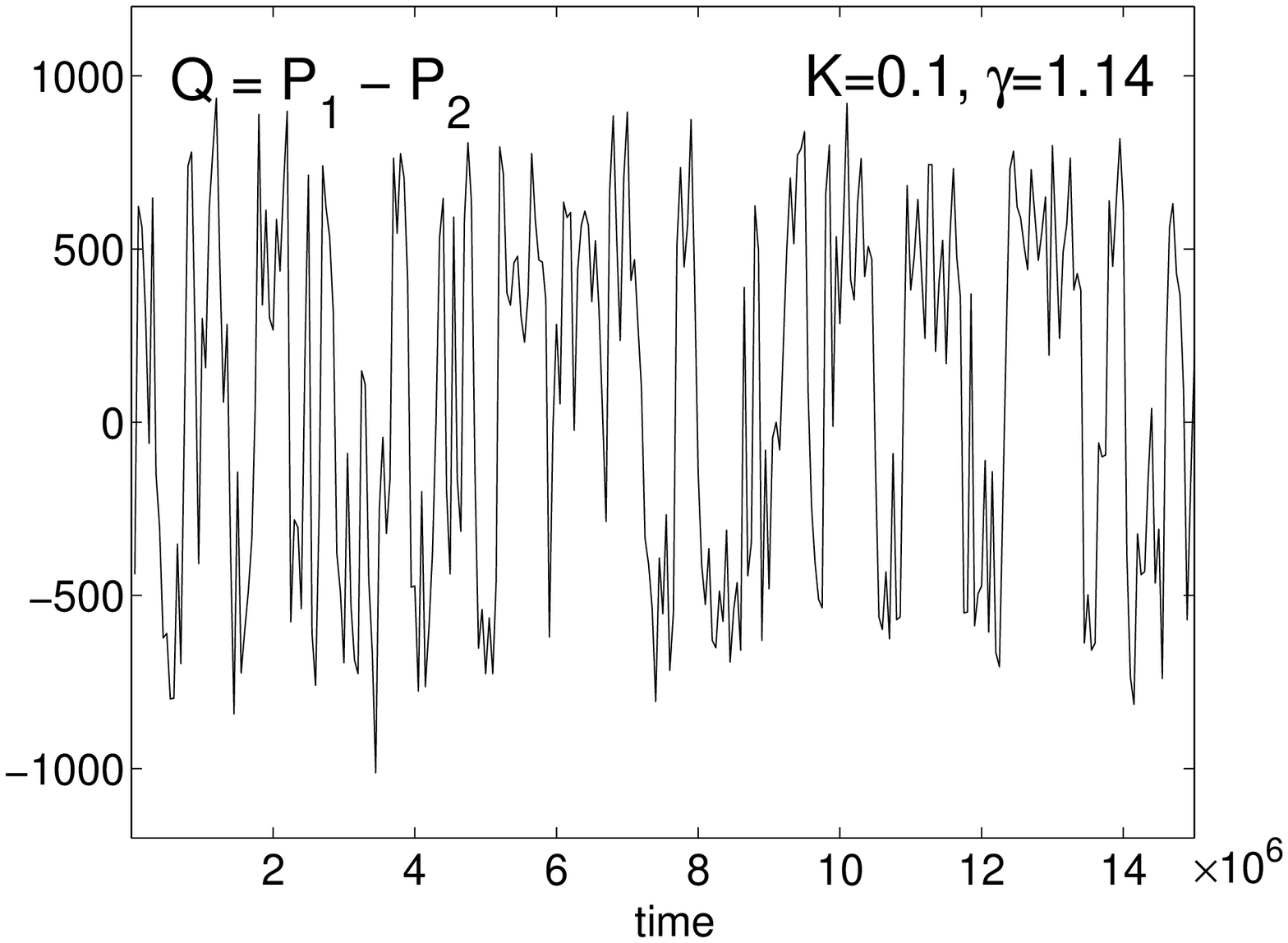,height=2.1in}
}
\bigskip
\centerline{{\large $K= 1$, $\gamma = 1.14$}}
\smallskip
\centerline{
\psfig{file=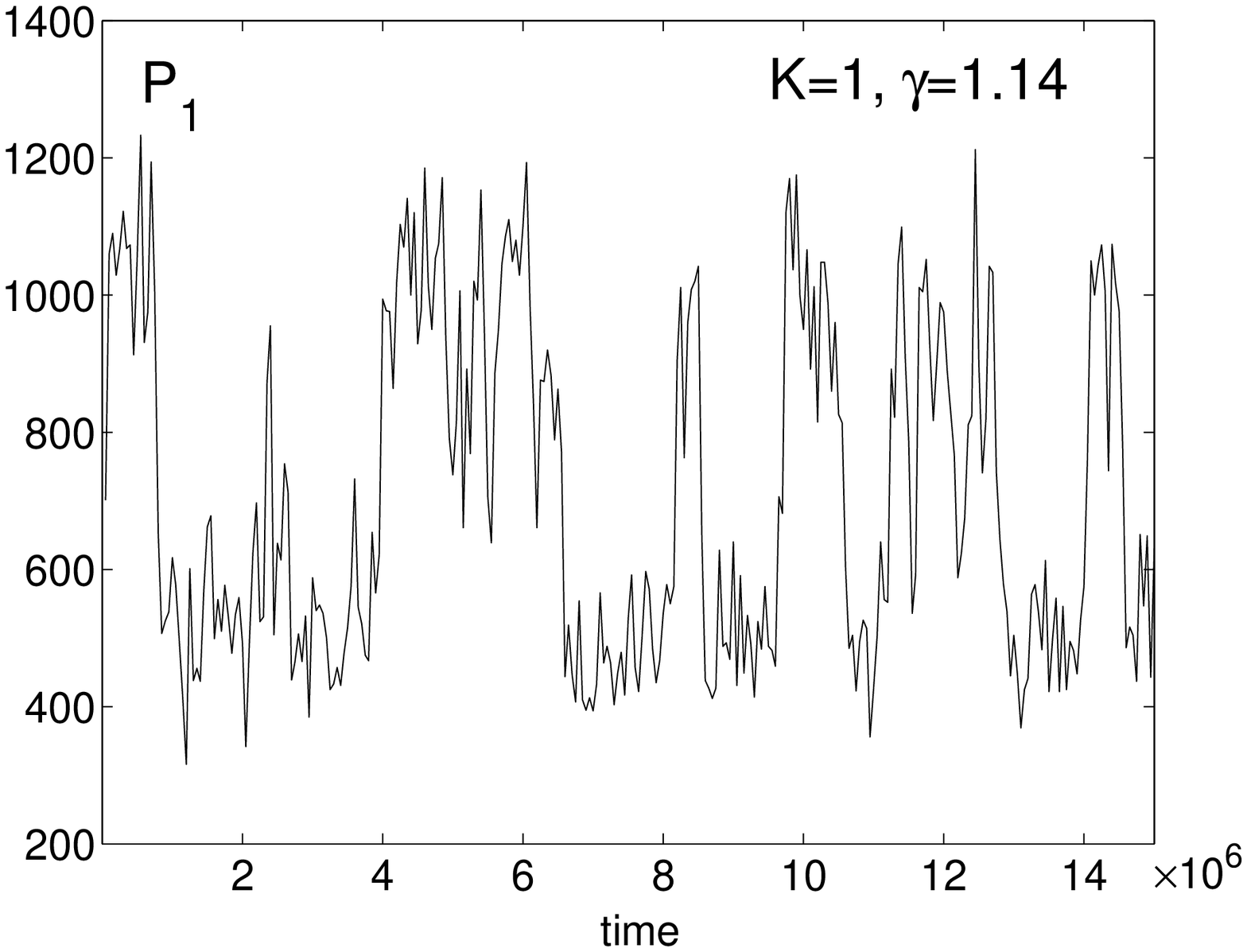,height=2.1in}
\quad
\psfig{file=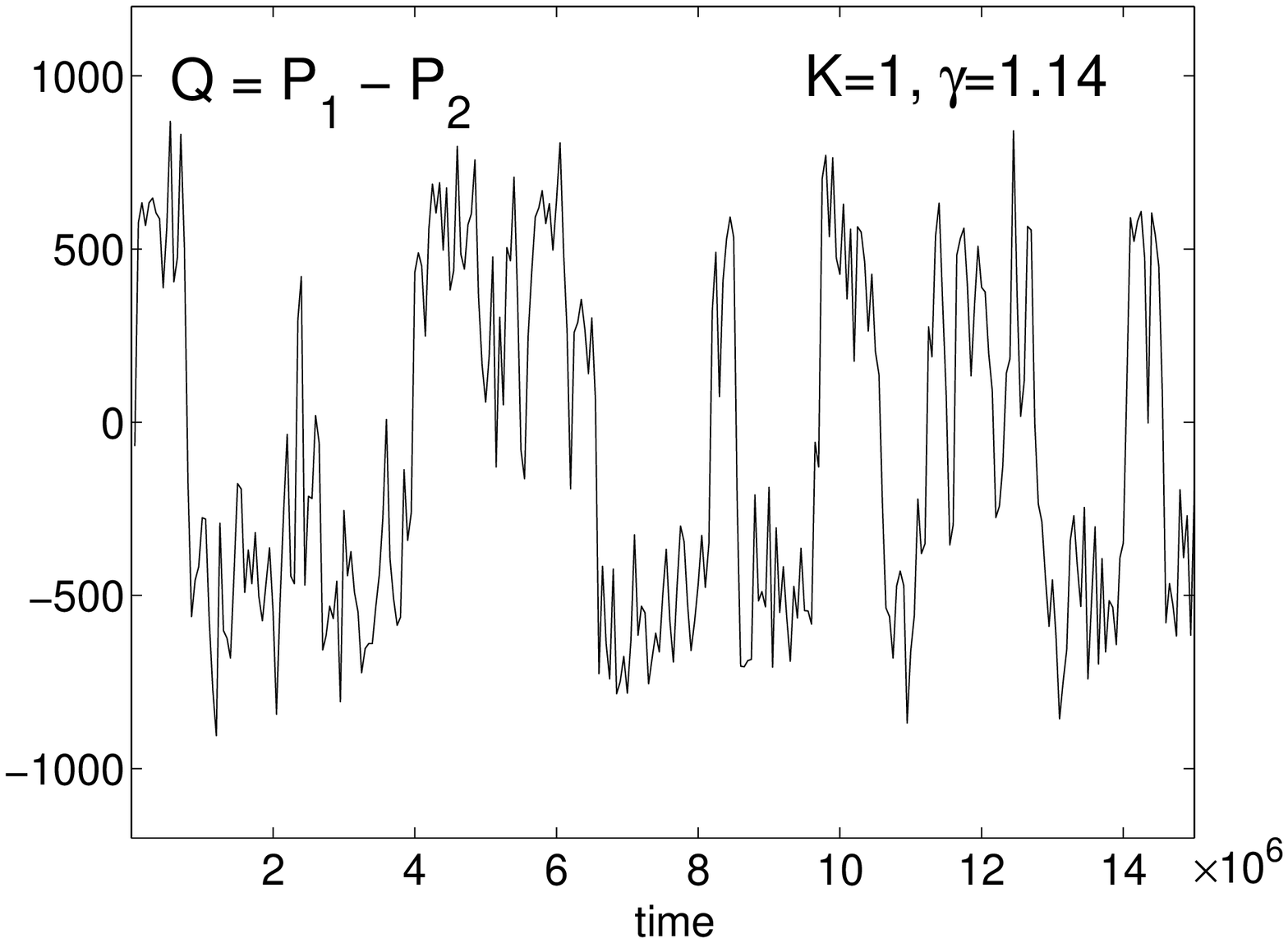,height=2.1in}
}
\bigskip
\centerline{{\large $K= 10$, $\gamma = 1.14$}}
\smallskip
\centerline{
\psfig{file=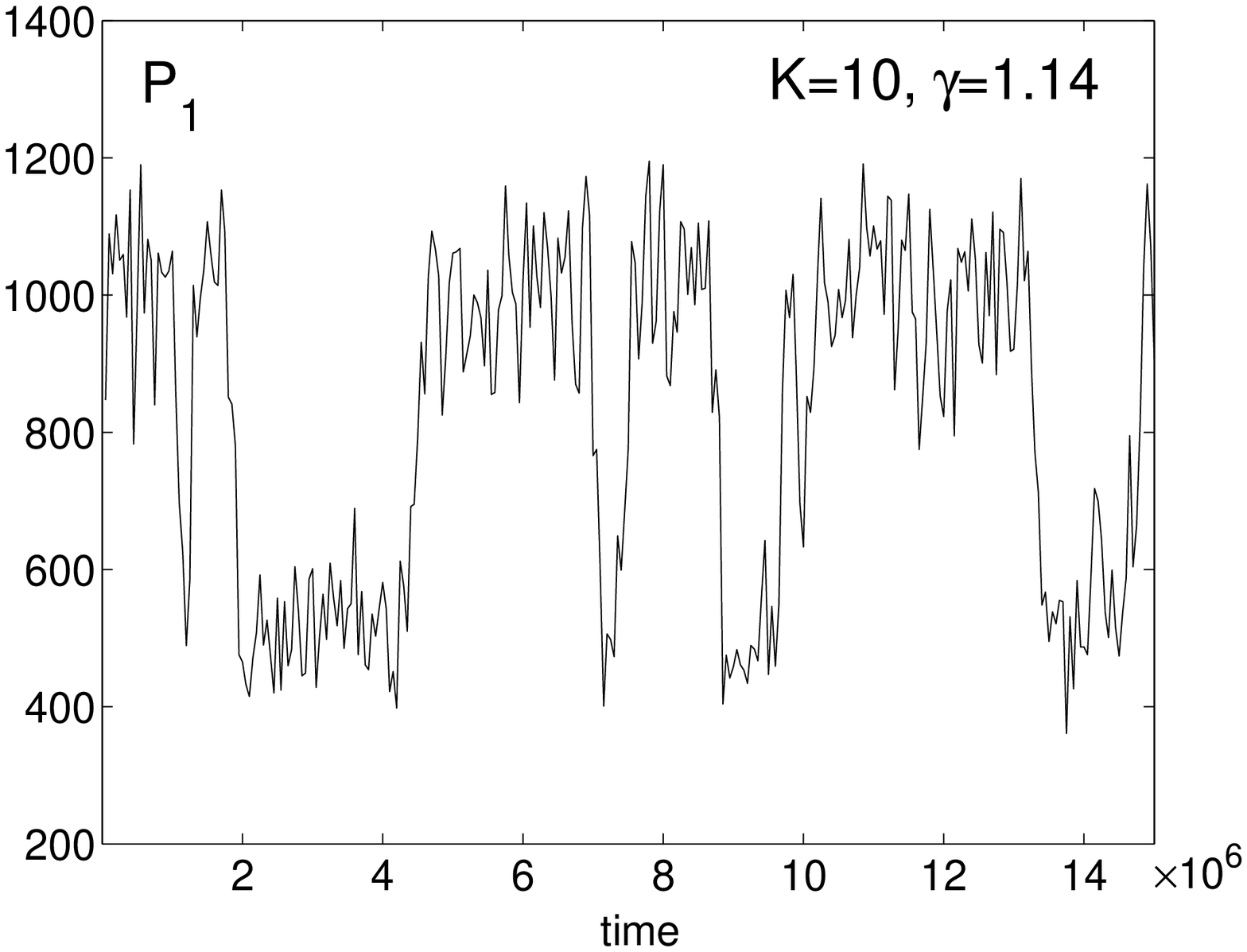,height=2.1in}
\quad
\psfig{file=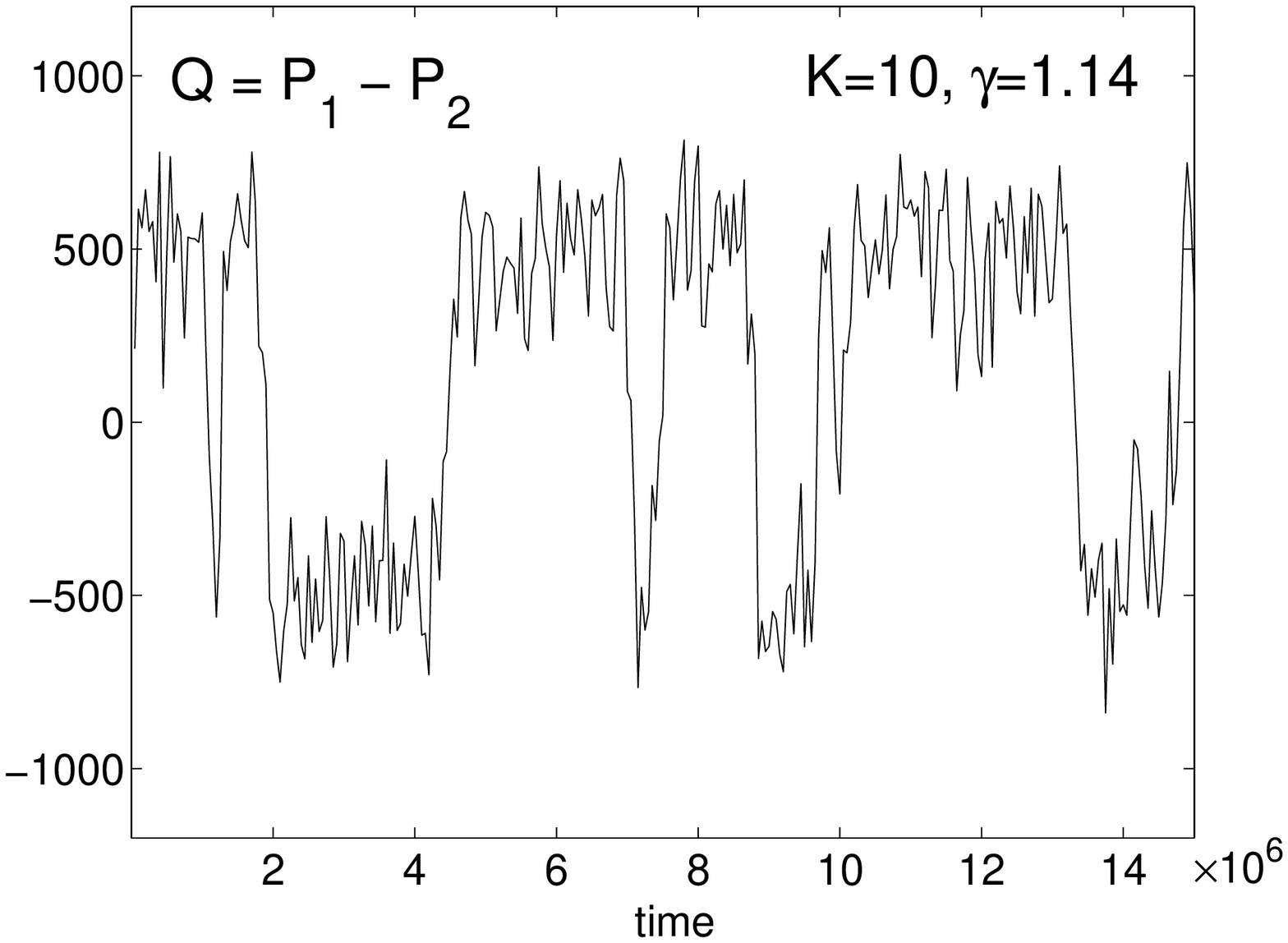,height=2.1in}
}
\caption{{\it
Stochastic Model II. Plots of $P_1$ and $Q=P_1 - P_2$ as a function
of time for different values of $K$ and $\gamma = 1.14.$
The other model parameter values are the same as in Figure $\ref{figPQ1}$.}}
\label{figPQB}
\end{figure}

Again, we use the standard Gillespie SSA \cite{Gillespie:1977:ESS} to simulate
model (\ref{prodP1b}) -- (\ref{prodO2b}).
The results for different values of $K$
for $\gamma=1.14$ are plotted in Figure
\ref{figPQB}. 
Comparing Figure \ref{figPQB}
and corresponding panel from Figure \ref{figPQ1}, we can confirm
that Stochastic Model II produces the same behavior as Stochastic Model I for large $K.$
However, in general, different values of $K$ can change the bifurcation 
structure of the system and affect the first
passage times between the two stable steady states
of the bistable system \cite{Kepler:2001:STR}. 

Because
Stochastic Models I and II do not explicitly take into account dimerization,
which in general is a fast process, they run much more efficiently than the full model
given by (\ref{prodP1}) -- (\ref{onoffO2}).
However, they do not in general preserve the noise 
structure of the full system. 
In the
next section we use all three models to highlight the computational 
features (and potential benefits) of equation-free
analysis.   
  
\section{Results of Equation-Free Analysis}

\label{secnumer}

In our approach, we want to study the stochastic
models presented above using only short bursts of appropriately
initialized stochastic simulations; the goal is to design
these bursts, and process their results so as to 
determine long-time properties of the system (e.g., steady-state
distributions, bifurcations, mean first passage times)
efficiently.
We use (and compare) the different algorithms
discussed in Section \ref{secmathfram}.

\subsection{The effective potential and steady state distribution}

\label{secsteadistr}

In this section, we use equation-free analysis to evaluate the effective
potential (an ``effective free energy") 
and the steady-state distribution for Stochastic Models I and II
and the full system. 
We start with Stochastic Model I. First, we will consider
the slow variable $Q \equiv P_1 - P_2$ and the fast (slaved) 
variable $R \equiv P_1 + P_2.$
Initially the preparatory step (A) of the algorithm presented in 
Section \ref{secmathfram} was done using the
method outlined in (A1), i.e. we used the conditional mean 
$<\!\!R|Q=q\!\!>$ to initialize the computations in the step (B). 
A good approximation to this average can be found 
using the deterministic equations, and this was the number used
in our preliminary computations to initialize the simulations
in the step (B). 
That is, for a given $Q$, we initialized all realizations 
in the step (B) with the same value of $R$. 
Then we chose
$\Delta t$ equal to 100 time steps of the Gillespie SSA. 
Note that
this implies that the actual value of $\Delta t$ varies for each realization and
depends on the values of the rate constants. 
However, the computer (CPU) time is the same for all the results presented for this
case. 

\begin{figure}
\centerline{
\psfig{file=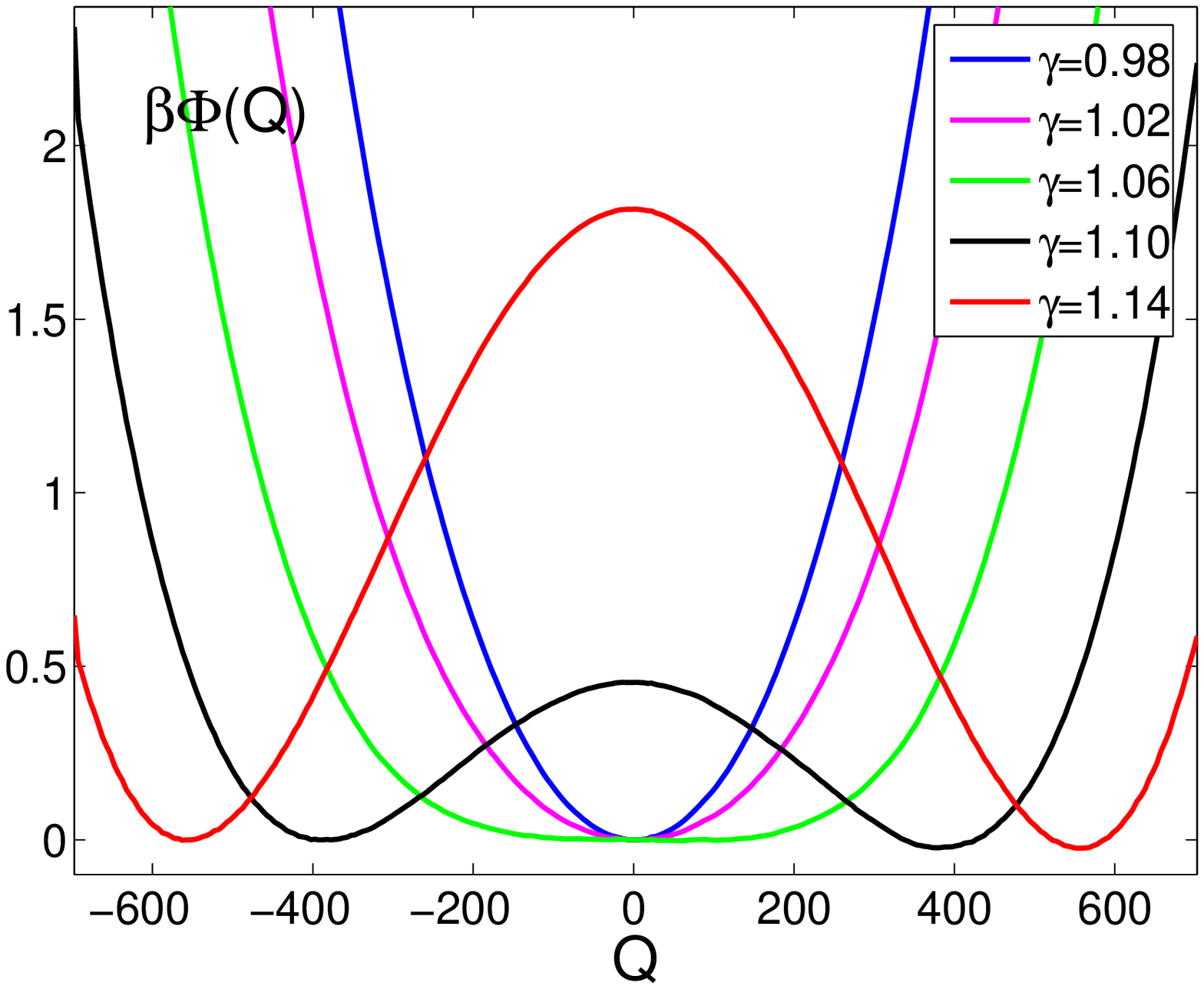,height=2.6in}
\psfig{file=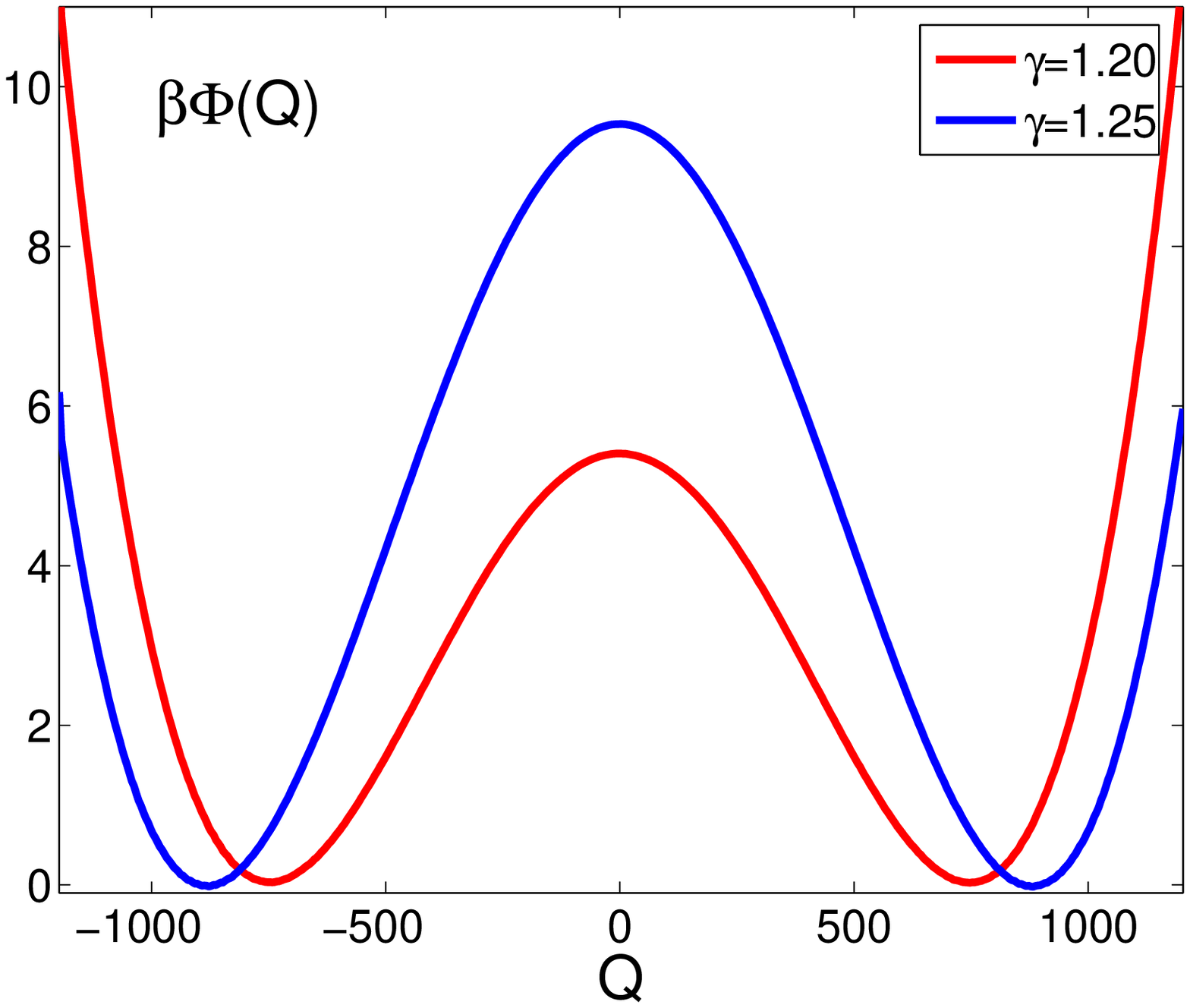,height=2.6in}
}
\caption{{\it The effective free energy $\Phi$ for different values of $\gamma$
computed by our procedure. 
The other model parameter values 
are the same as in Figure $\ref{figPQ1}$.}}
\label{PhiQsymmetric}
\end{figure}

The equation-free results for the effective potential for different values of 
$\gamma$ are given in Figure \ref{PhiQsymmetric}.
These results are in good agreement with the long-term
stochastic simulations presented in 
Section \ref{secmodelA}. 
The potential has a single minimum $\gamma < 1.06$. As
$\gamma$ is increased the potential broadens implying the system 
becomes ``noisier".
When $\gamma > 1.06$,
the potential shows two local minima and the system is bistable.

\begin{figure}[ht]
\centerline{
\psfig{file=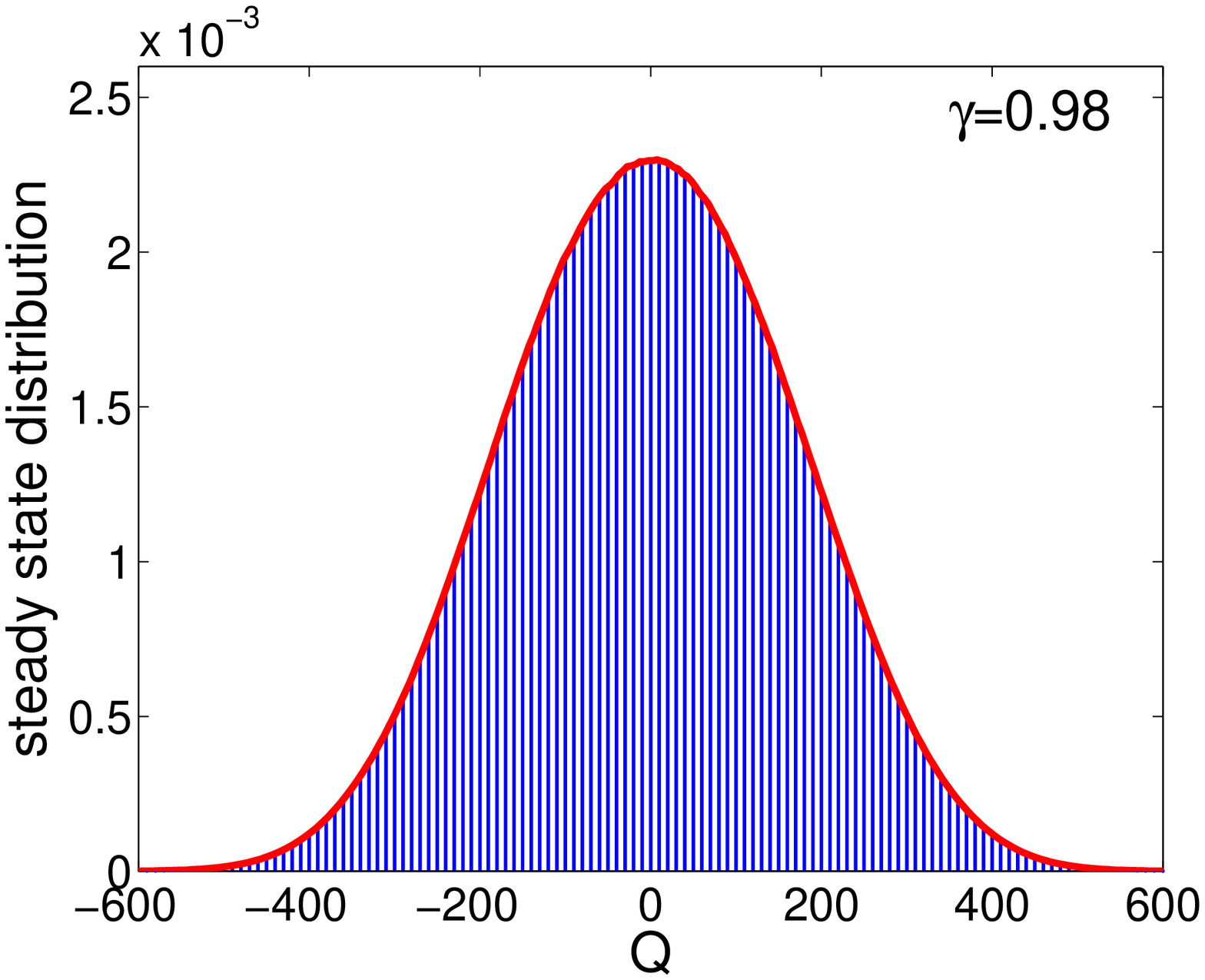,height=2.3in}
\psfig{file=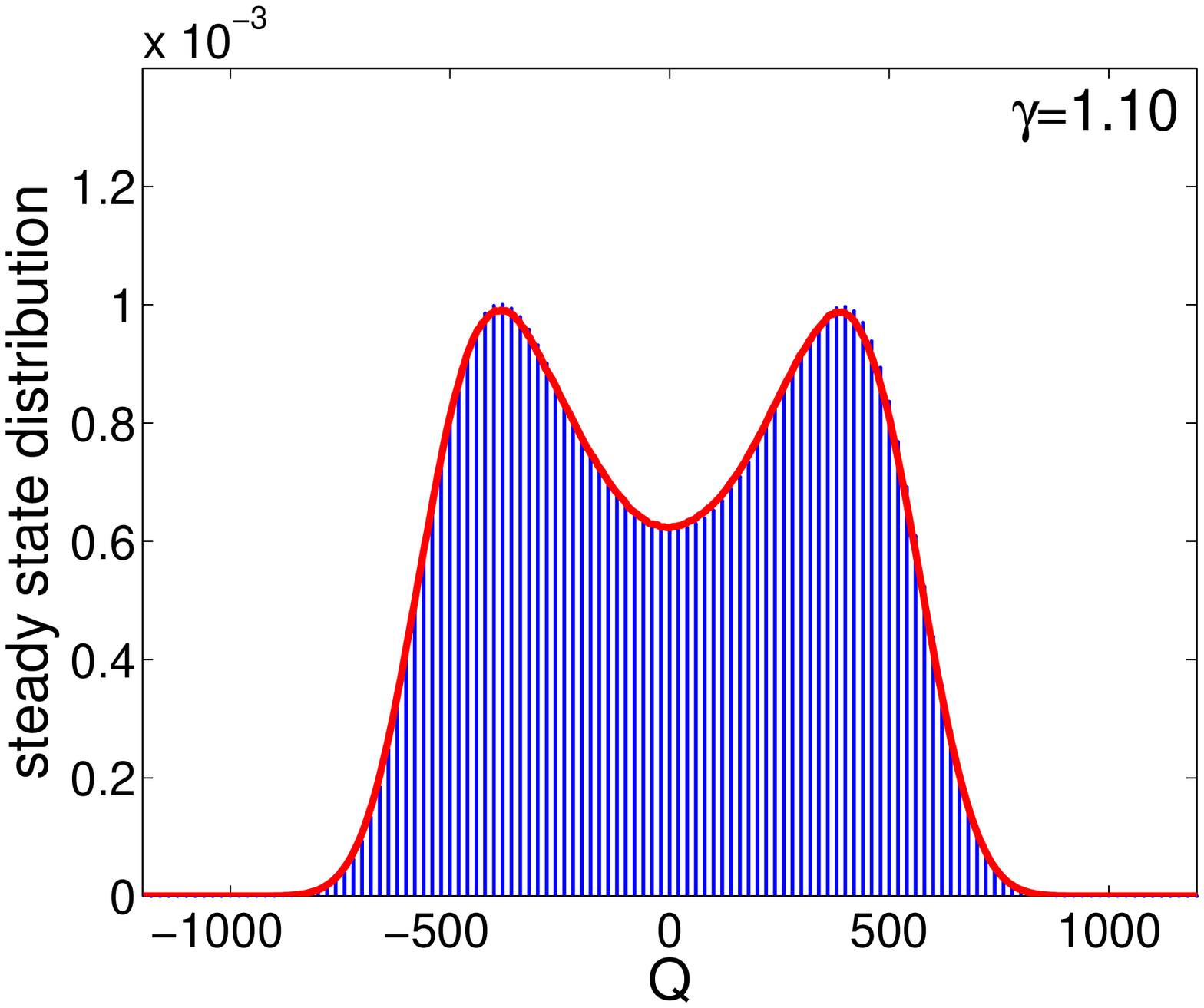,height=2.3in}
}
\centerline{
\psfig{file=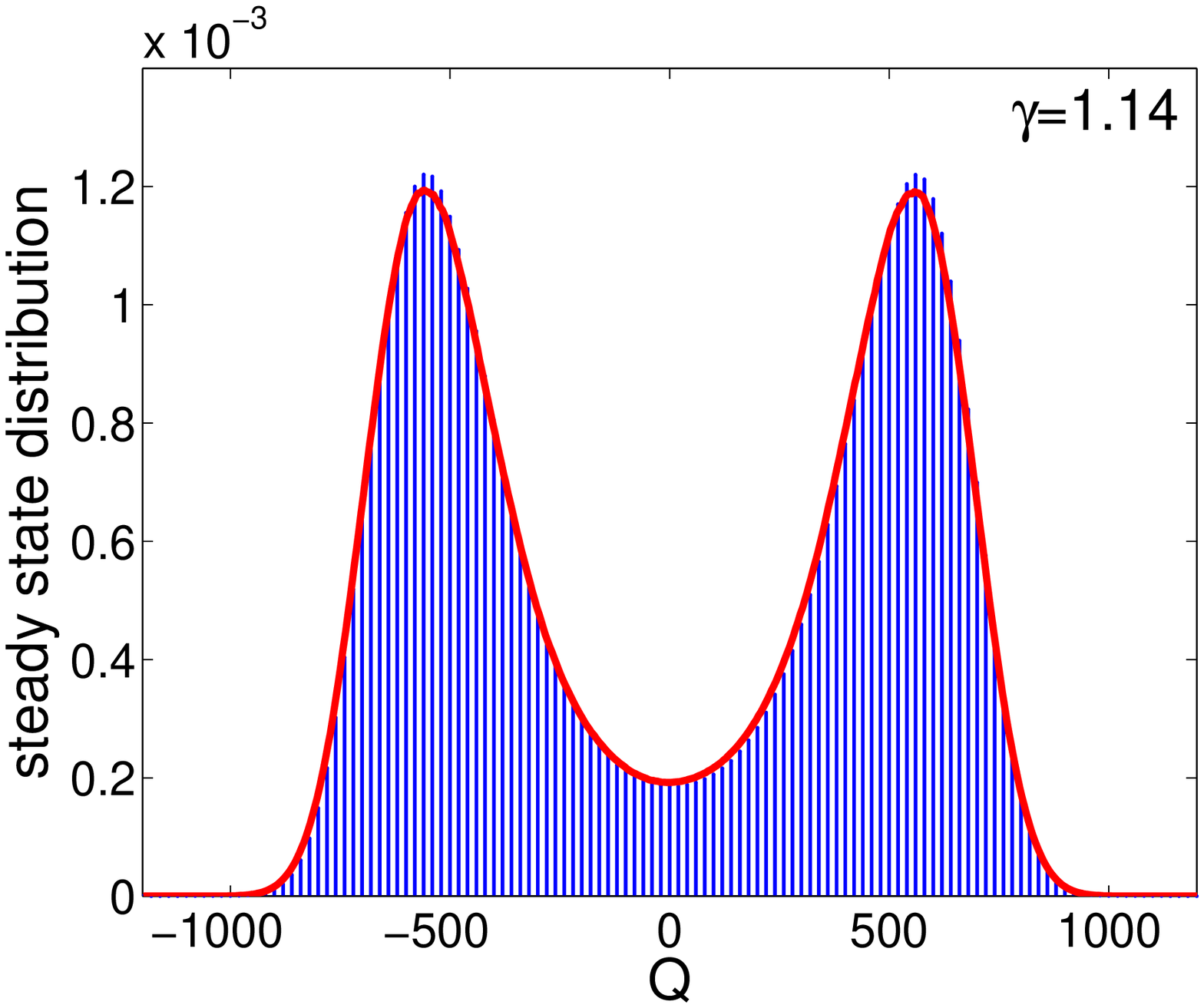,height=2.3in}
\psfig{file=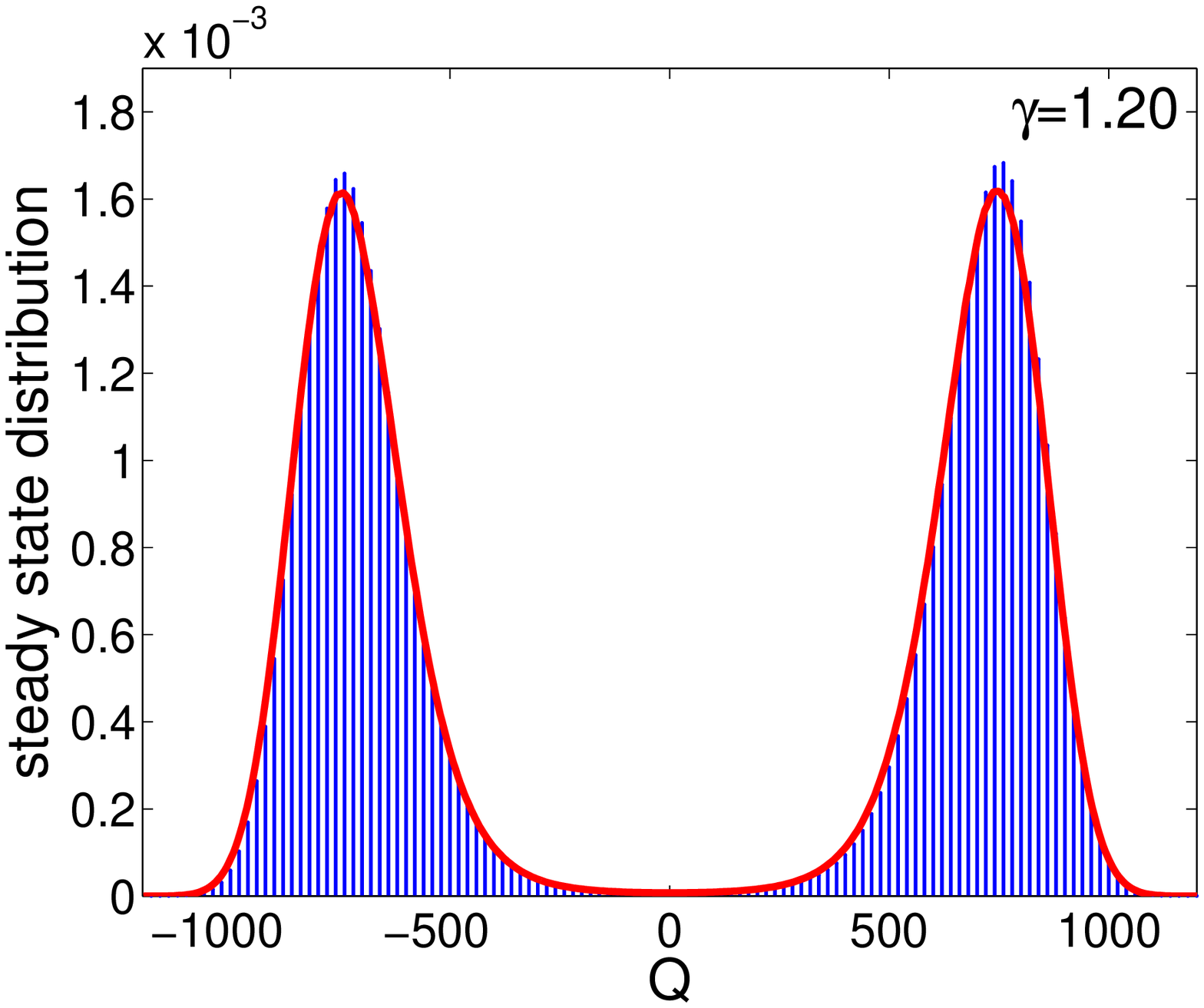,height=2.3in}
}
\caption{{\it Comparison of steady state distributions obtained from the
effective free energies shown in Figure $\ref{PhiQsymmetric}$ 
with histograms (blue) obtained by long time
simulations.}}
\label{PhiQcomphist}
\end{figure}

Since we are using a very simple stochastic model, it is not
computationally expensive to compute the steady-state distributions
directly by long time simulations. 
We use the Gillespie SSA to generate $10^{11}$
time steps of the stochastic process and recorded the value
of $Q$ at each time step. 
The resulting time series was binned to produce
the steady-state
distribution of the system. 
Figure \ref{PhiQcomphist} presents a comparison
of the two computed steady state distributions. 
The results
obtained by long-time simulations are shown as blue histograms and
the steady state distributions computed from the
effective potential $C \exp[- \beta \Phi(Q)]$  are given by the red lines.
We see that equation-free analysis gives very good results.

In Section \ref{secmathfram}, we introduced three possible
methods, (A1) -- (A3), to perform the preparatory step (A)
(typically called the ``lifting" step in the equation-free
framework).
We have shown
that approach (A1) produces good results for Stochastic Model 1.
Since (A1) works, there is no need to improve
the results by considering (A2). 
Instead, we discuss the (A3) approach. 
In this approach given $Q=q$ 
we run the simulations for a short time $\delta t$ and record the
value of $R$.
Then we reset $Q=q$ but leave $R$ unchanged.
We repeat this procedure many times  and compute the conditional density 
$P(r|Q=q)$ as a histogram of recorded values of $R$. 
In our simulations, we chose $\delta t$ equal to one SSA step. 
To compute the conditional density $P(r|Q=q)$, 
we used 11 million SSA steps.
First we let the system run for a  million time steps to remove the transient
in $R$, and then used remaining 10 million time steps to compute the conditional
density. 
We used 200 million Gillespie time steps in part (B) of the algorithm.
Consequently,  step (A3) did not significantly 
change the computational cost of the program. 

\begin{figure}[ht]
\centerline{
\psfig{file=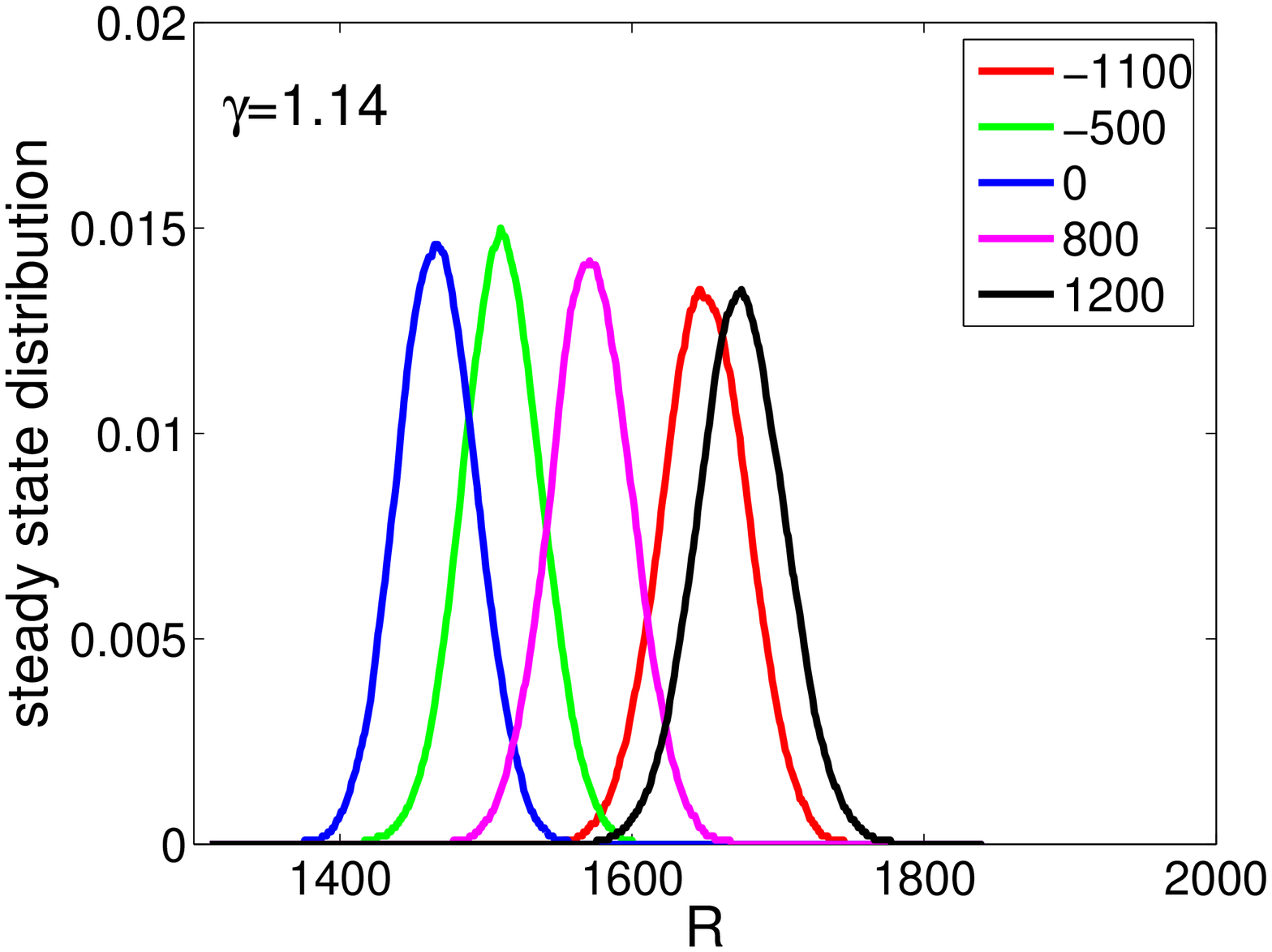,height=2.3in}
\psfig{file=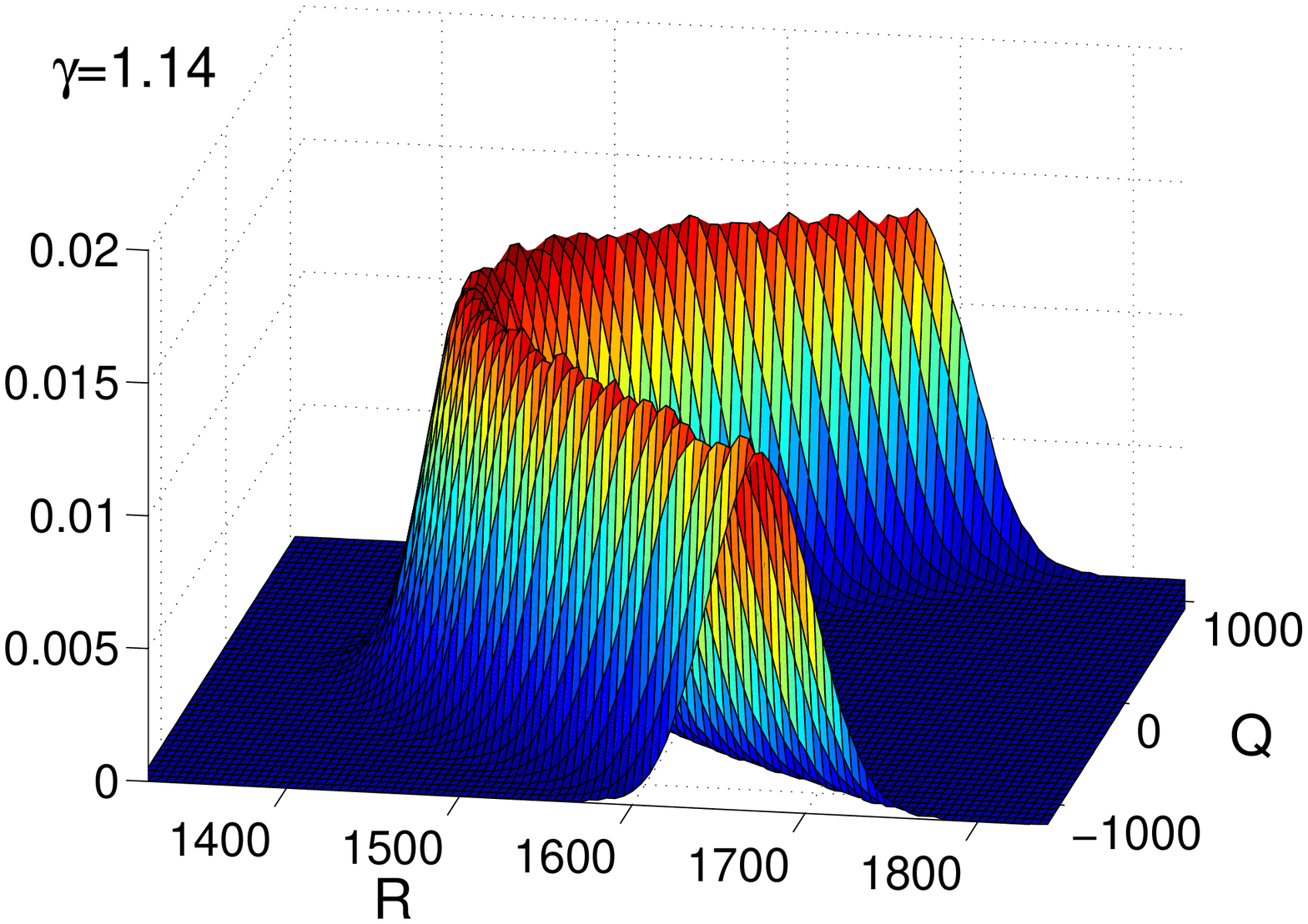,height=2.3in}
}
\centerline{
\psfig{file=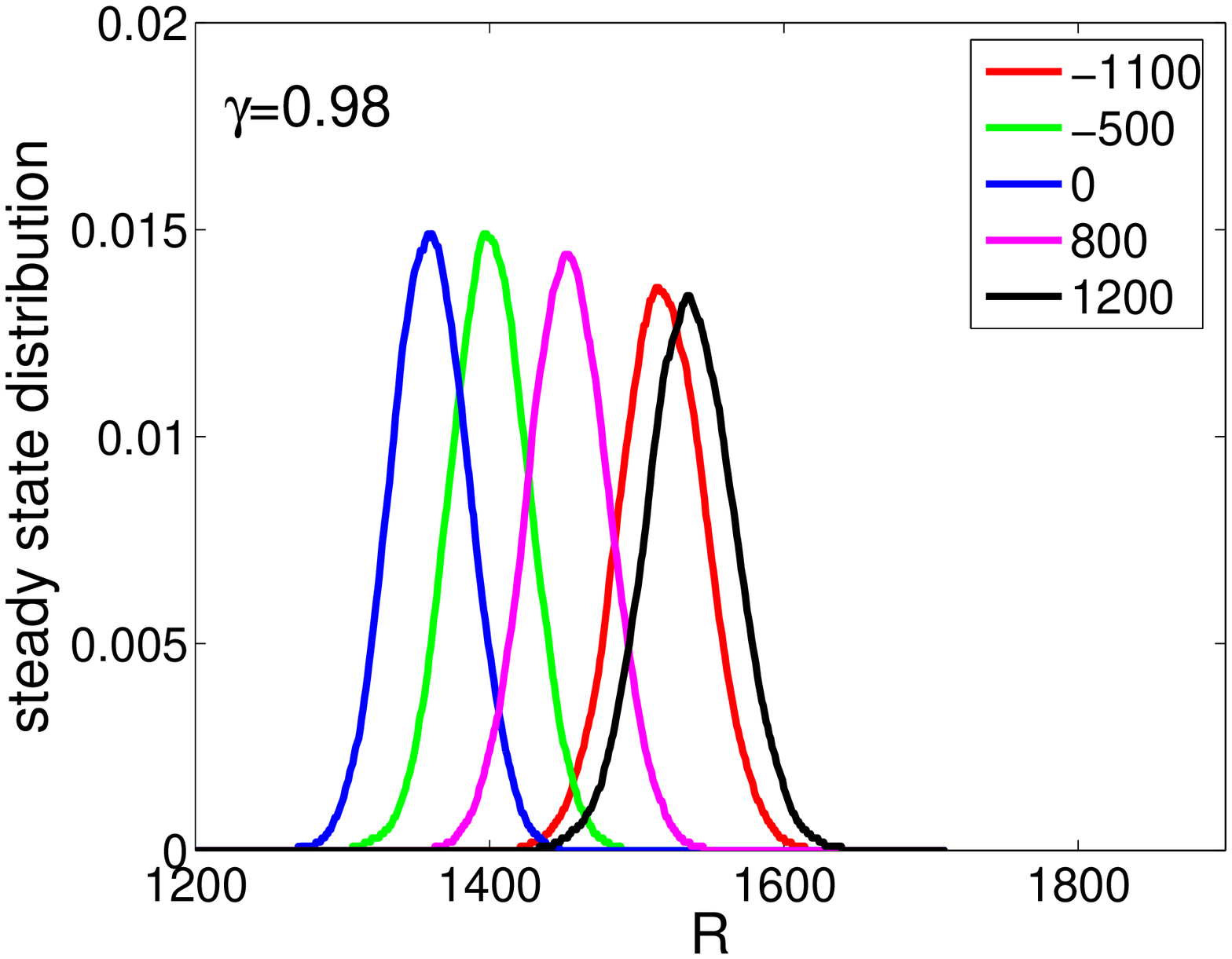,height=2.3in}
\psfig{file=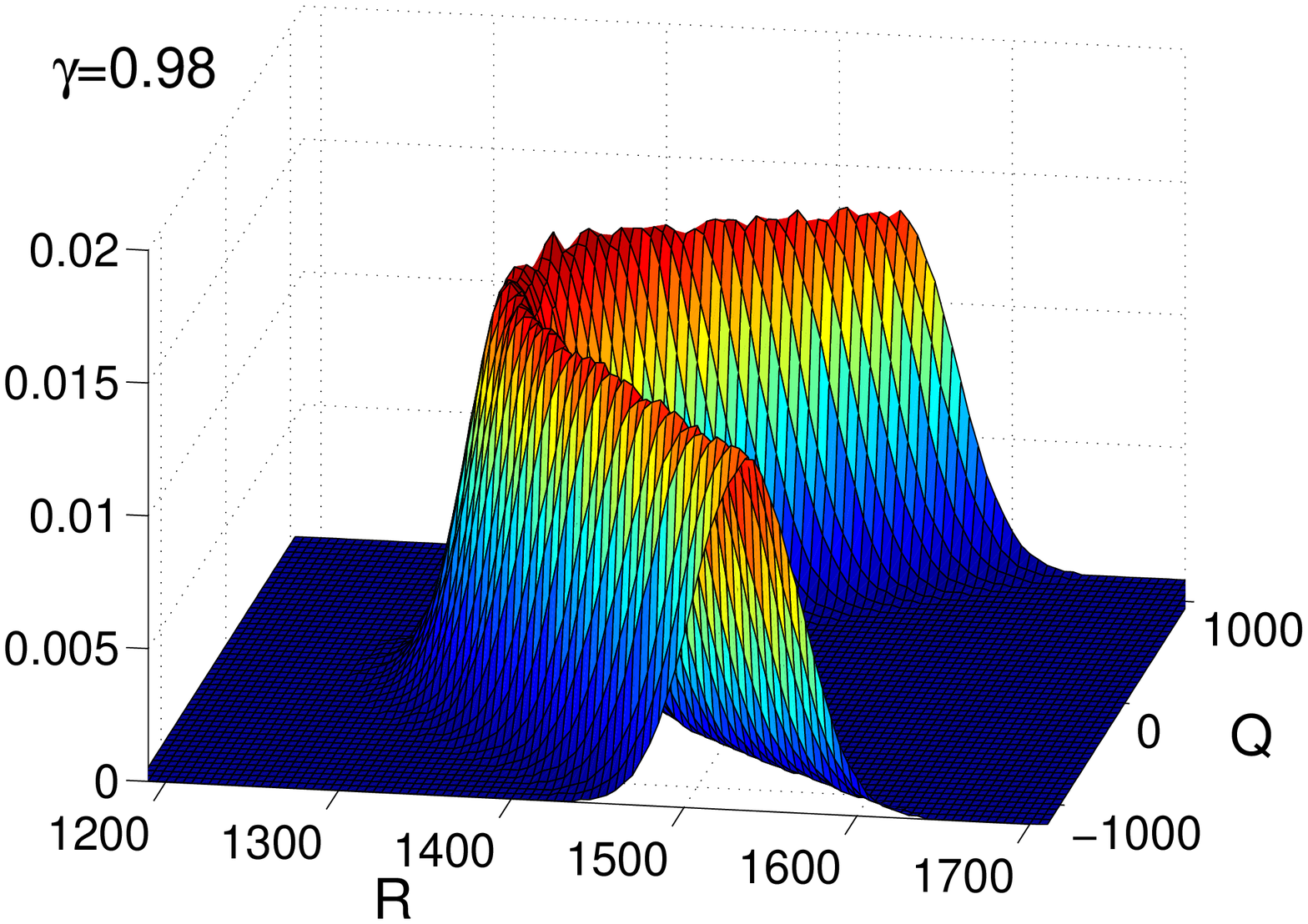,height=2.3in}
}
\caption{{\it Conditional distribution $P(r|Q=q)$ for Stochastic
Model I. Pictures on the left show  $P(r|Q=q)$ for selected
values of $q$. Pictures on the right show $P(r|Q=q)$ as a function
of $r$ and $q$.}}
\label{PhiRcon}
\end{figure}

The graphs of $P(r|Q=q)$ for $\gamma = 0.98$ and $\gamma = 1.14$
are given in Figure \ref{PhiRcon}. 
The left panel in these figures shows 
$P(r|Q=q)$ for five values of $Q$. 
The right panels show  $P(r|Q=q)$ as a function of $r$
and $q$. 
Next, we can use the computed conditional density $P(r|Q=q)$ to initialize
$R$ in the step (B). 
Doing this, produces results which
are virtually identical to results from Figure \ref{PhiQcomphist}
(graphs not shown).

\begin{figure}[ht]
\centerline{
\psfig{file=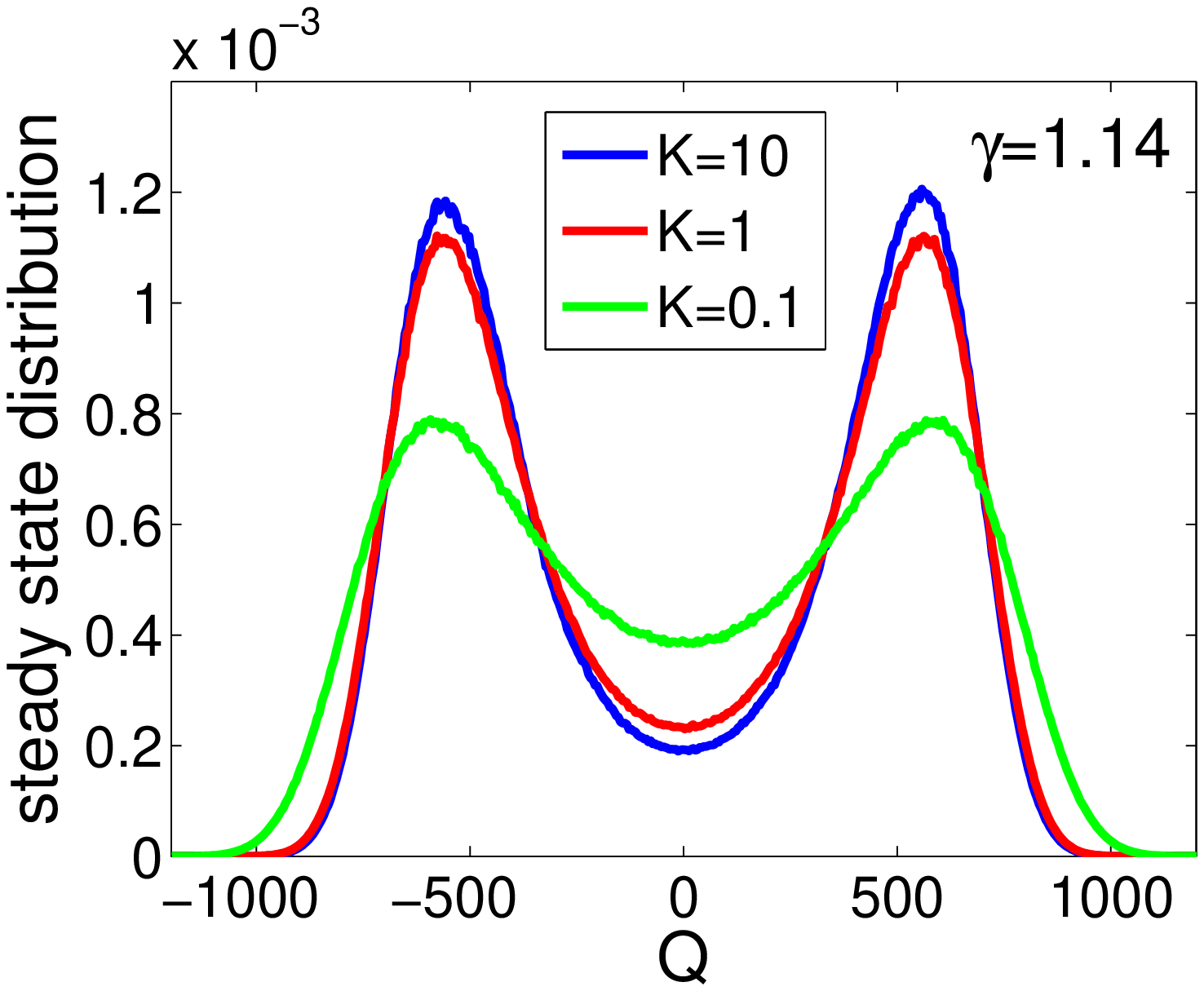,height=2.3in}
\psfig{file=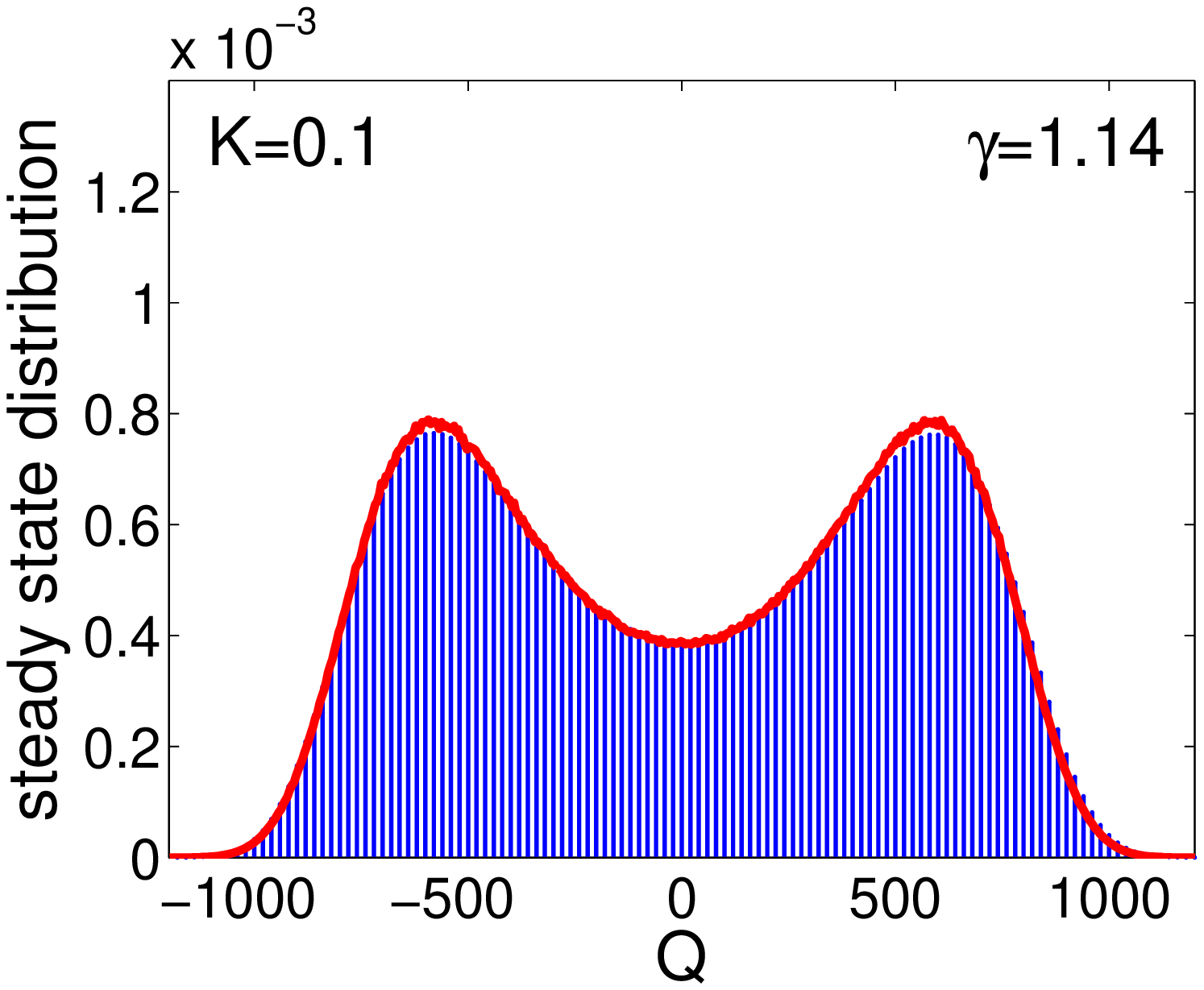,height=2.3in}
}
\centerline{
\psfig{file=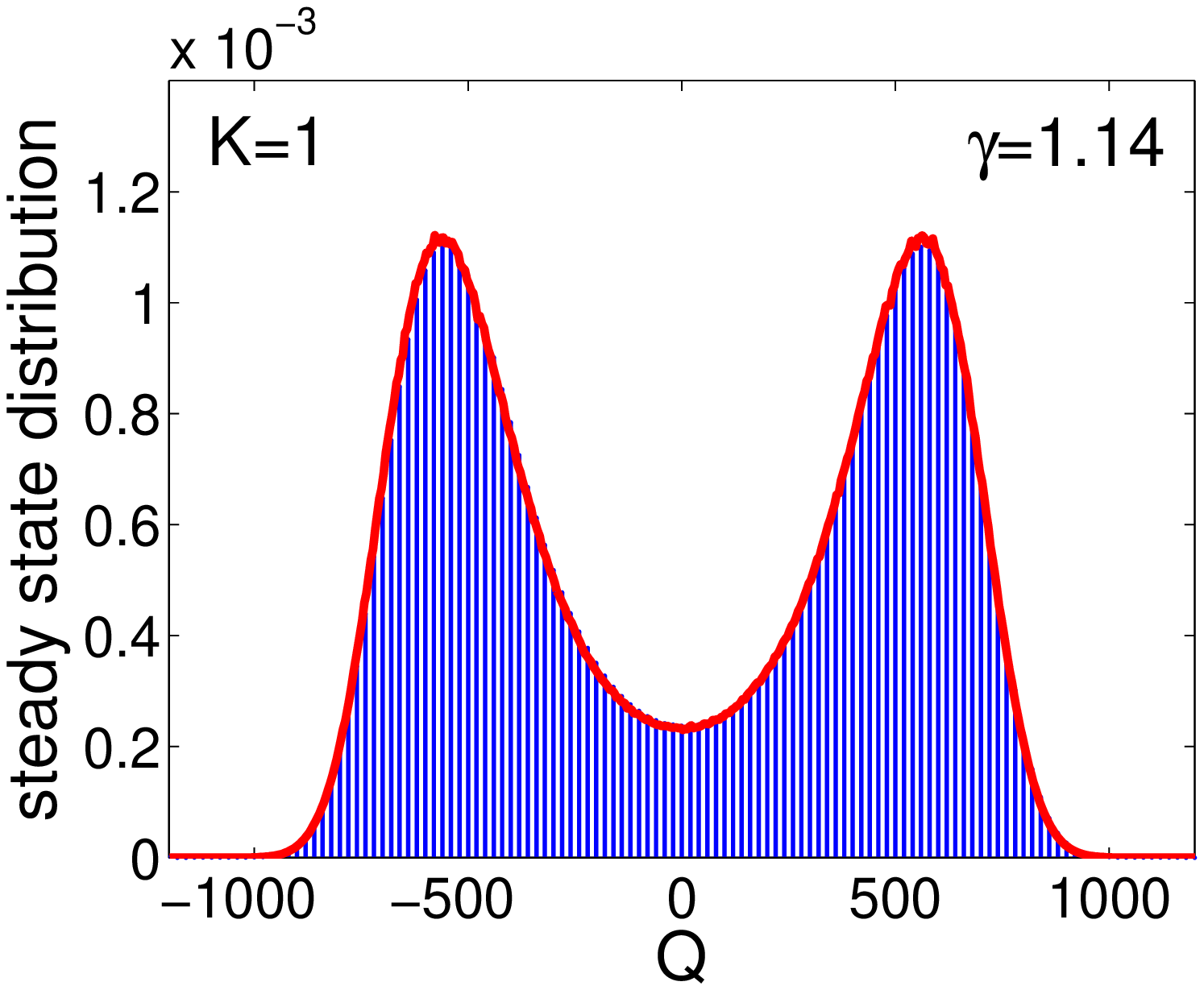,height=2.3in}
\psfig{file=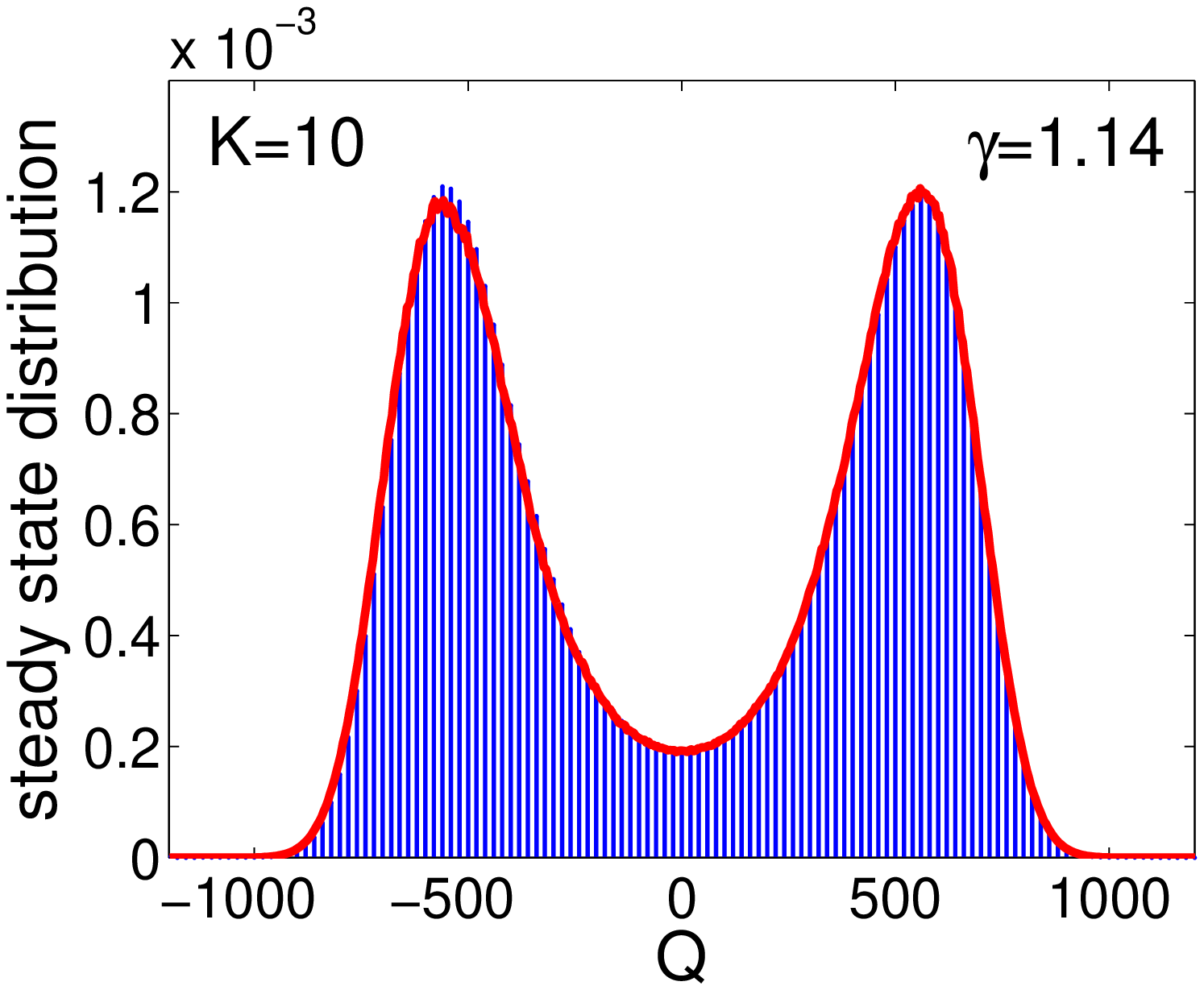,height=2.3in}
}
\caption{{\it Comparison of steady state distributions from 
Stochastic Model II. The top left panel
are results from the equation-free analysis for $K=0.1$, $K=1$ and $K=10$
and $\gamma = 1.14$. 
The remaining three panels compare these results (red lines) to the
steady state distribution computed from long-time Monte Carlo simulation (blue histograms).
The other model parameter values 
are the same as in Figure $\ref{figPQ1}$.}}
\label{PhiFM1}
\end{figure}

\begin{figure}[ht]
\centerline{
\psfig{file=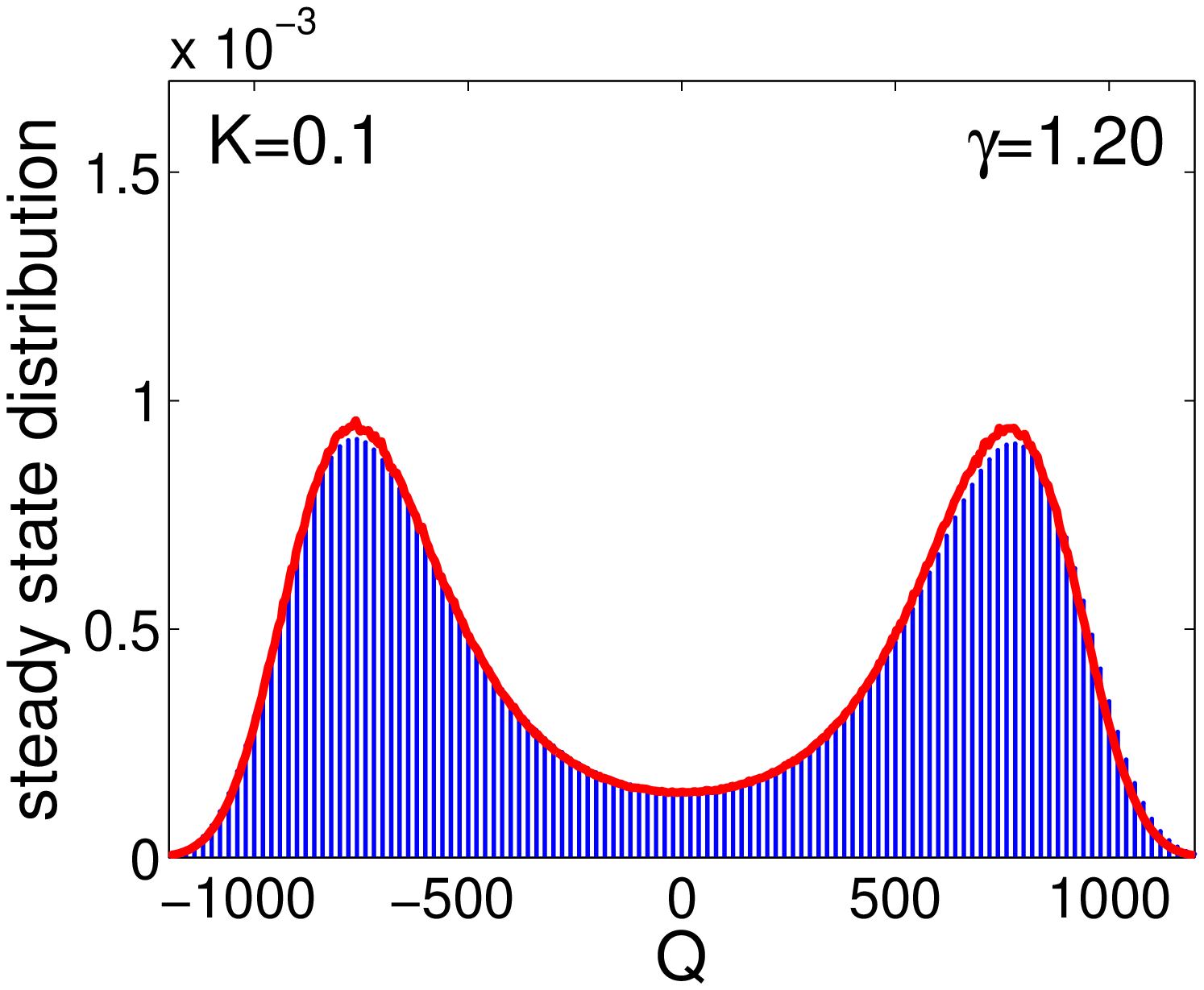,height=2.3in}
\psfig{file=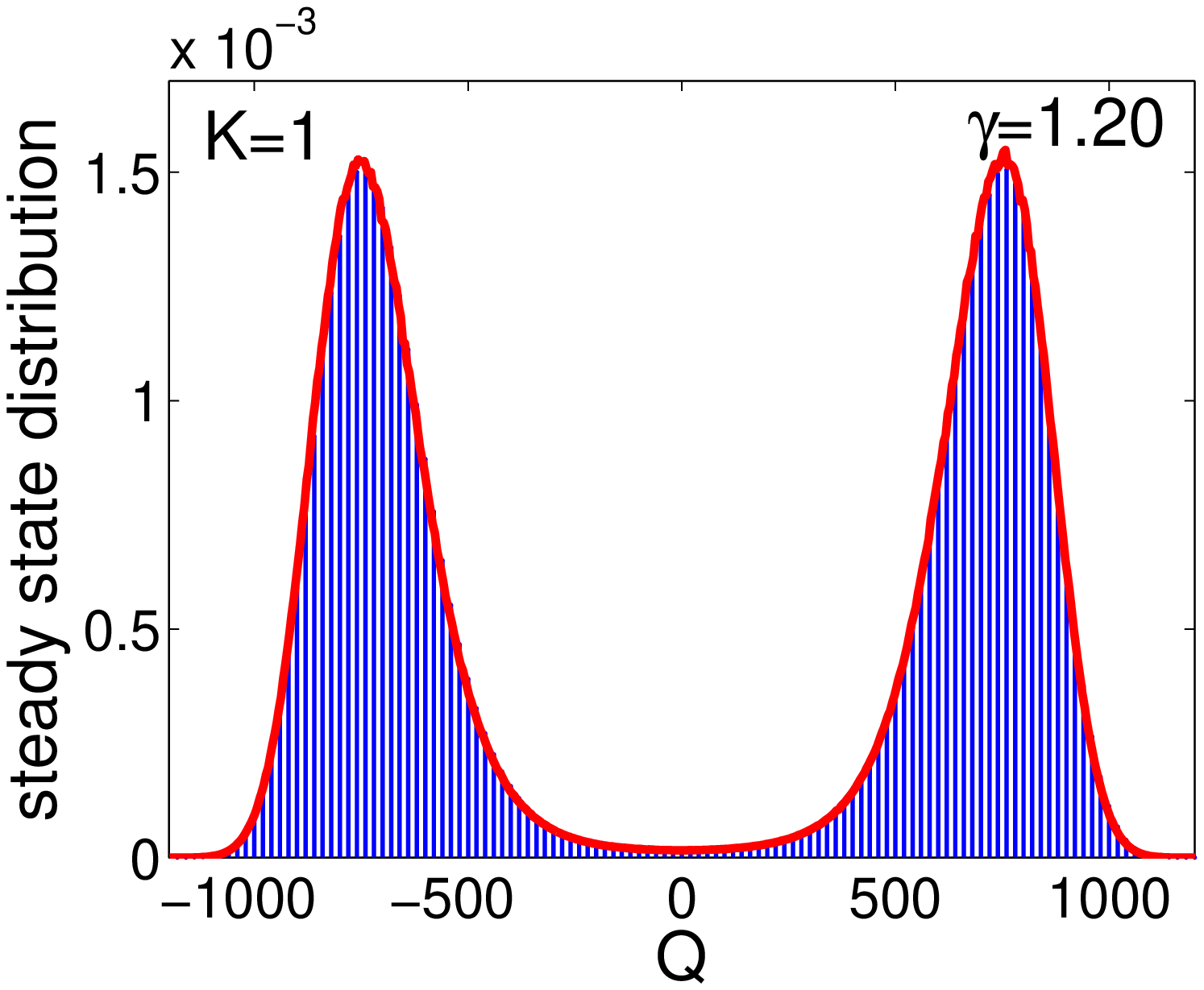,height=2.3in}
}
\caption{{\it Stochastic Model II. Comparison of steady state distributions 
obtained by  equation-free analysis (red line) 
with histograms (blue) obtained by long time
stochastic simulations. In this figure $\gamma = 1.2$ and $K= 0.1$ in 
the left panel and $K=1$ in the right. The other model parameter values 
are the same as in Figure $\ref{figPQ1}$.}}
\label{PhiFM2}
\end{figure}

We now repeat the previous computations using the more complicated 
Stochastic Model II. 
The results are shown in Figures \ref{PhiFM1} and \ref{PhiFM2}. 
In Figure \ref{PhiFM1}, we choose $\gamma = 1.14$ and compute the steady state 
distribution for $Q$ for three values of $K$. 
The results are compared
with direct simulations of Stochastic Model II and with each other.
The results from Figure \ref{PhiFM1} can also be compared 
to the corresponding plot with $\gamma = 1.14$ in Figure \ref{PhiQcomphist}, which 
can be viewed
as the limit $K \to \infty.$ 
As can be seen, the
results given by Stochastic Model II for $K=10$ are already 
in good agreement with the corresponding results obtained
by Stochastic Model I. 
Figure \ref{PhiFM2} shows similar results for
$\gamma = 1.20$. 
Again, we obtained
accurate results using the equation-free method.

\begin{figure}
\centerline{
\psfig{file=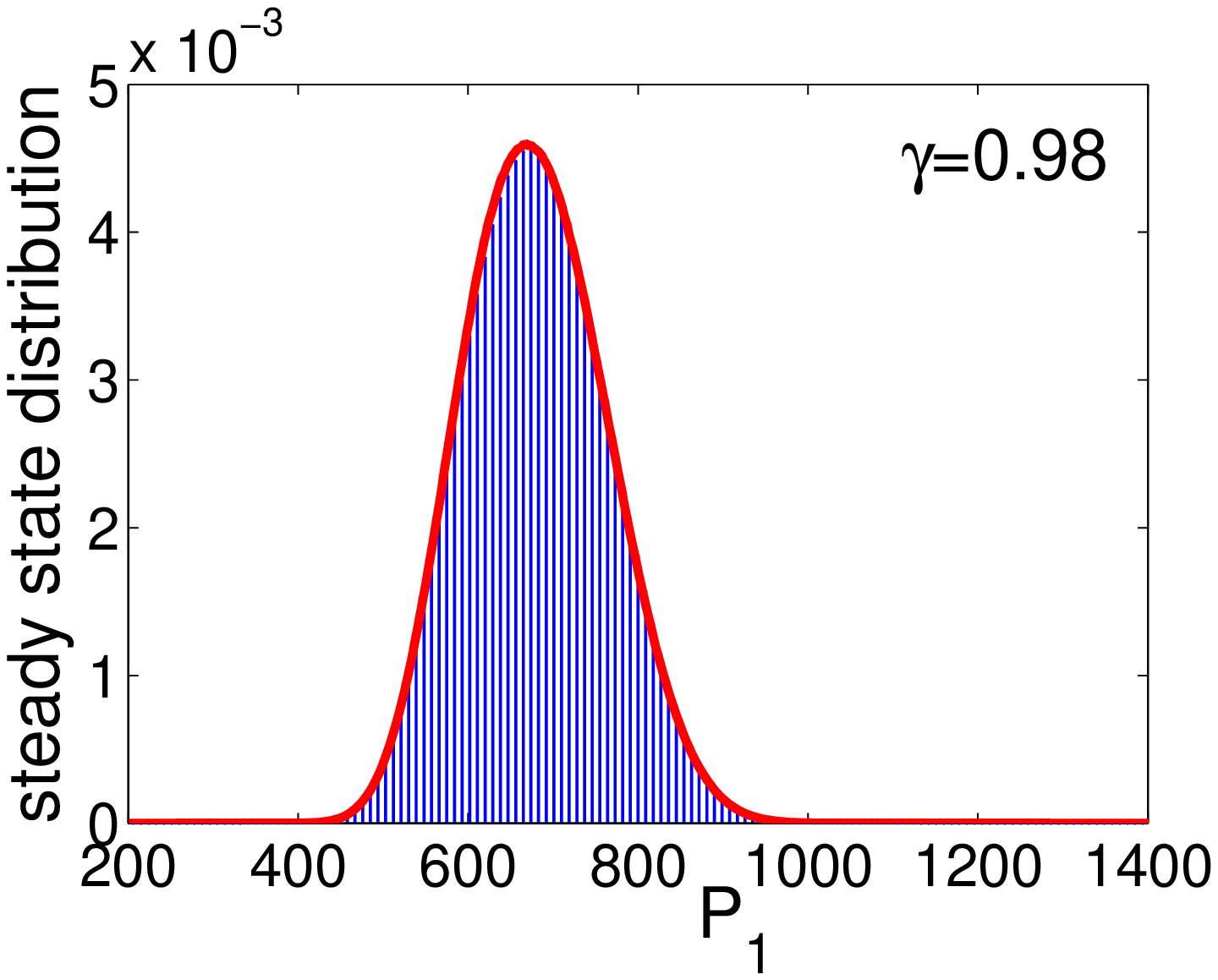,height=2.3in}
\psfig{file=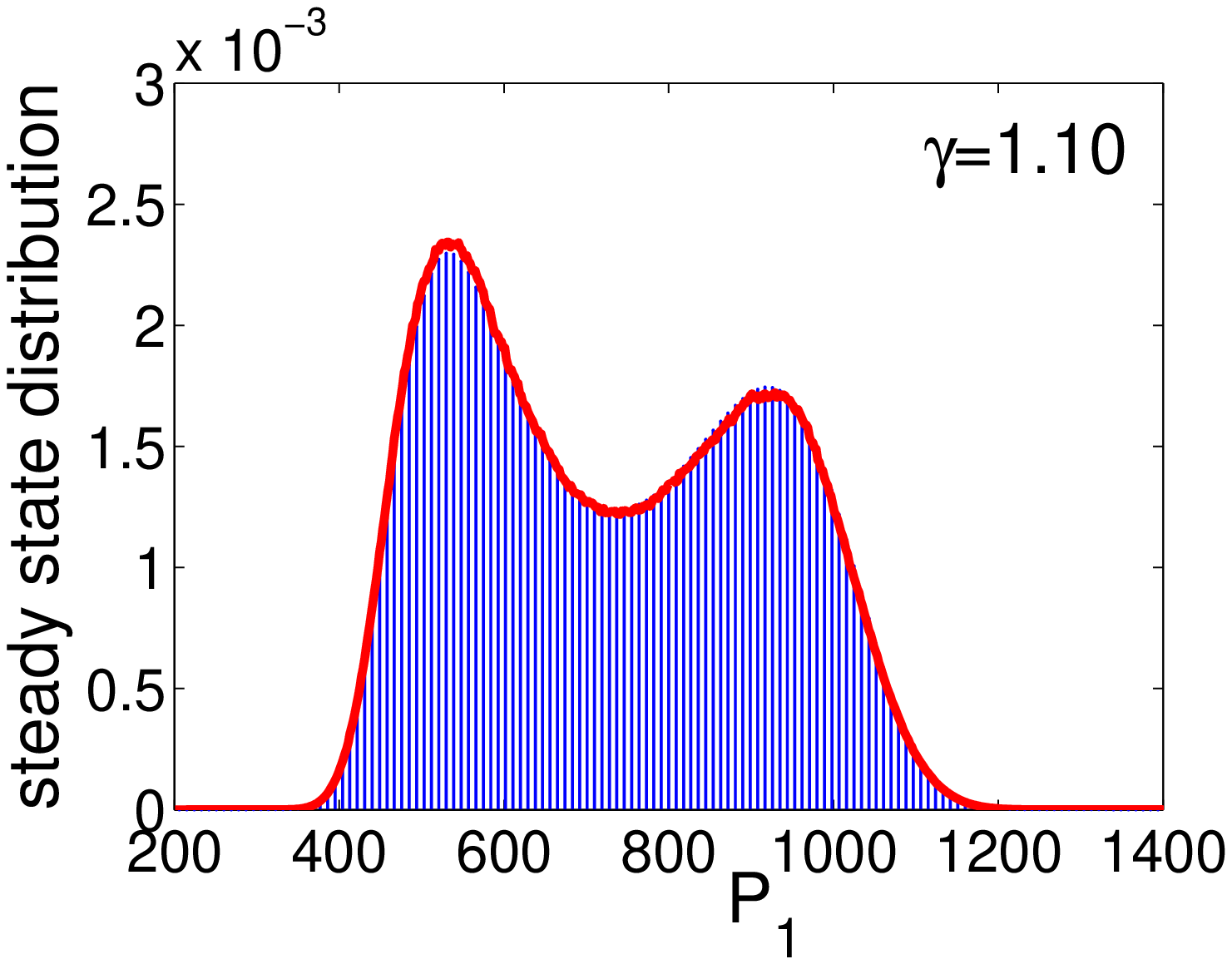,height=2.3in}
}
\centerline{
\psfig{file=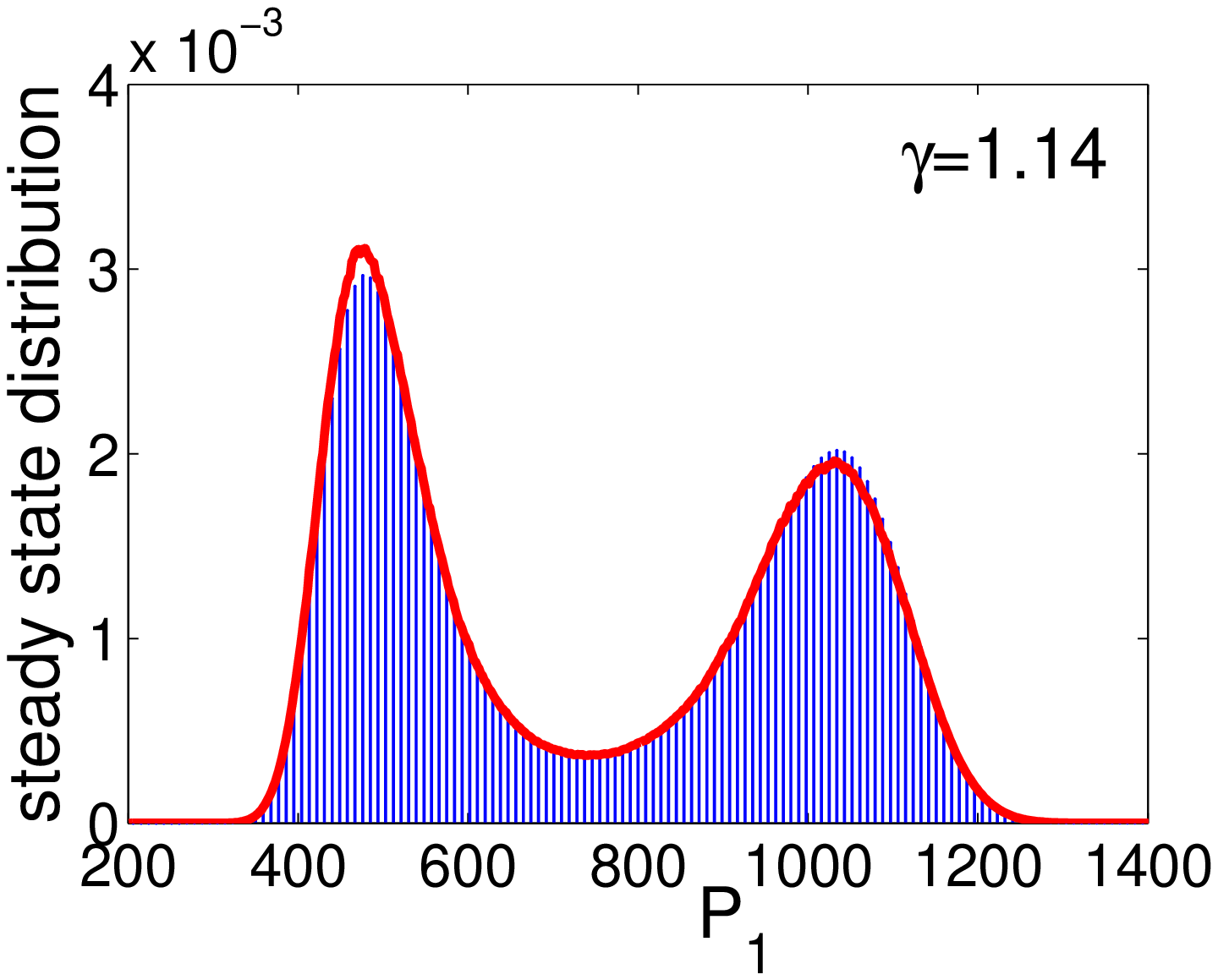,height=2.3in}
\psfig{file=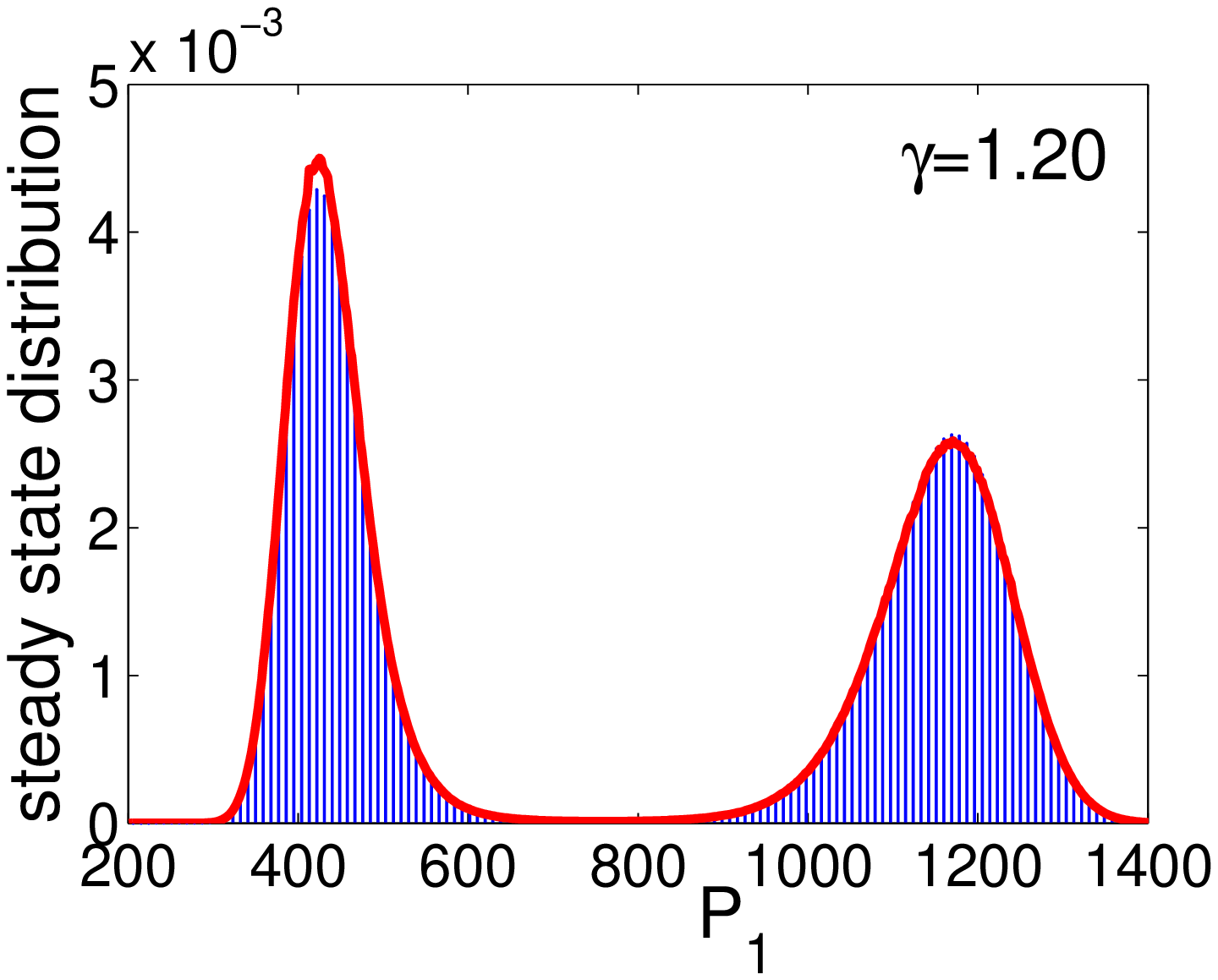,height=2.3in}
}
\caption{{\it Comparison of steady state distributions using
the variable $P_1$ as the observable for various values of
$\gamma$. The other model parameter values 
are the same as in Figure $\ref{figPQ1}$. 
Again, the red lines are the
results of equation-free analysis and the blue histograms are
obtained by the long-time stochastic simulations.}}
\label{figP1slow}
\end{figure}

Up to now we have used the symmetric variable $Q \equiv P_1 - P_2$ as our observable. 
However,  we often do not have {\it a priori} knowledge of the
slow variable, or, more generally, of a good observable to parametrize
the long-time system dynamics.
To investigate the sensitivity of our results
to the choice of observable, we repeated the computations on Stochastic
Model I  using $P_1$ instead of $Q$. 
To use $P_1$ as our observable,
we modify step (A1) so that we simply initialize $P_2$
using $P_2 = \frac{1}{\delta} \frac{\gamma}{1 + \omega P_1^2}.$
The numerical results for different values of $\gamma$ are given 
in Figure \ref{figP1slow}. 
Again good agreement is seen
between the equation-free method and the Monte-Carlo simulations. 
Because $P_1$ has both a slow and a fast component, this result 
illustrates that equation-free
methods can be used even when the slow variable is unknown.
An extensive discussion of this point in a deterministic context
can be found in \cite{Gear:2004:PSM}: one does not necessarily need 
{\it the correct}
slow variable -- one needs an observable that {\it parametrizes}
the slow manifold, a quantity in terms of which the slow manifold can be
expressed as the graph of a function.

\begin{figure}
\picturesAB{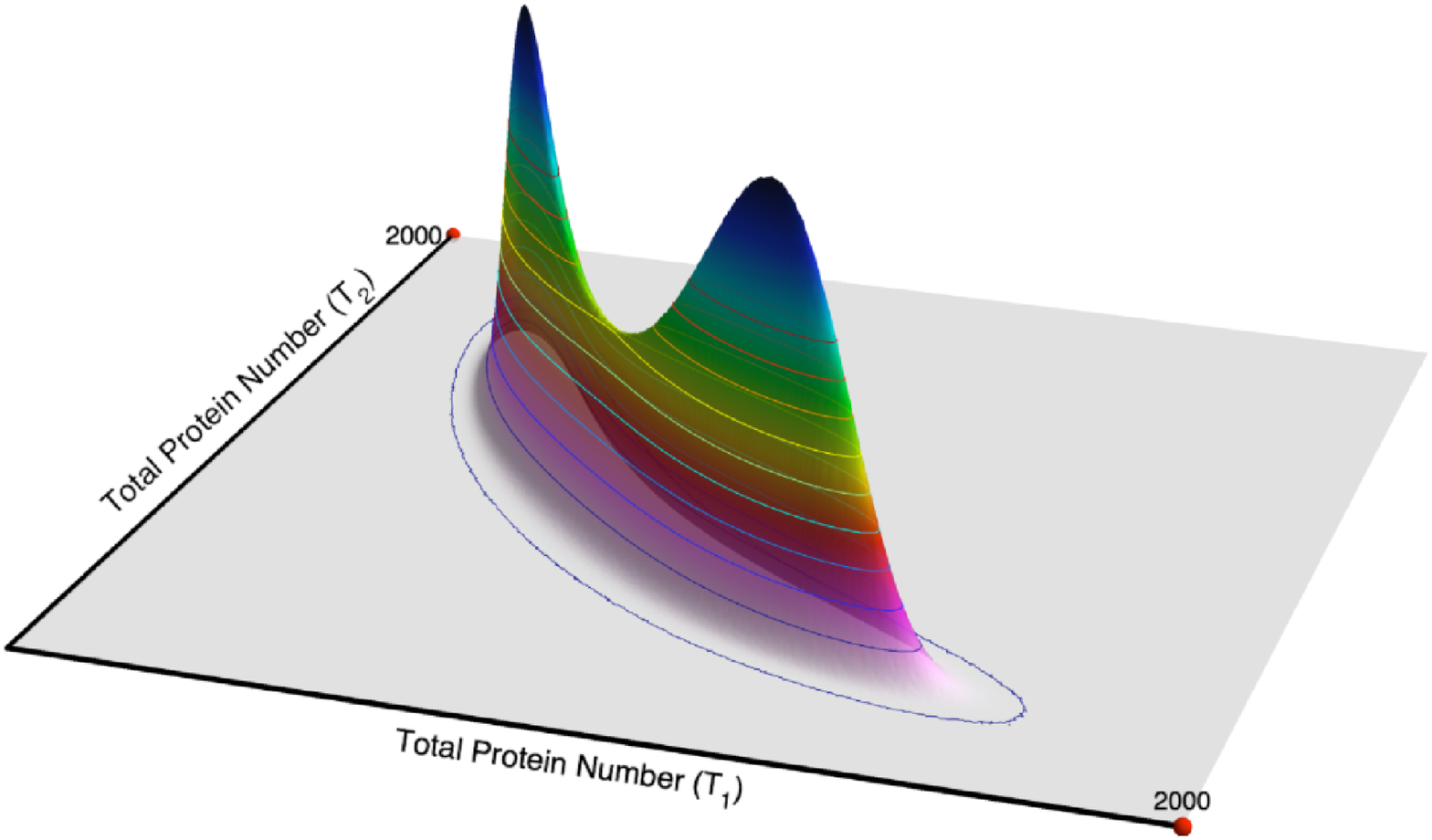}{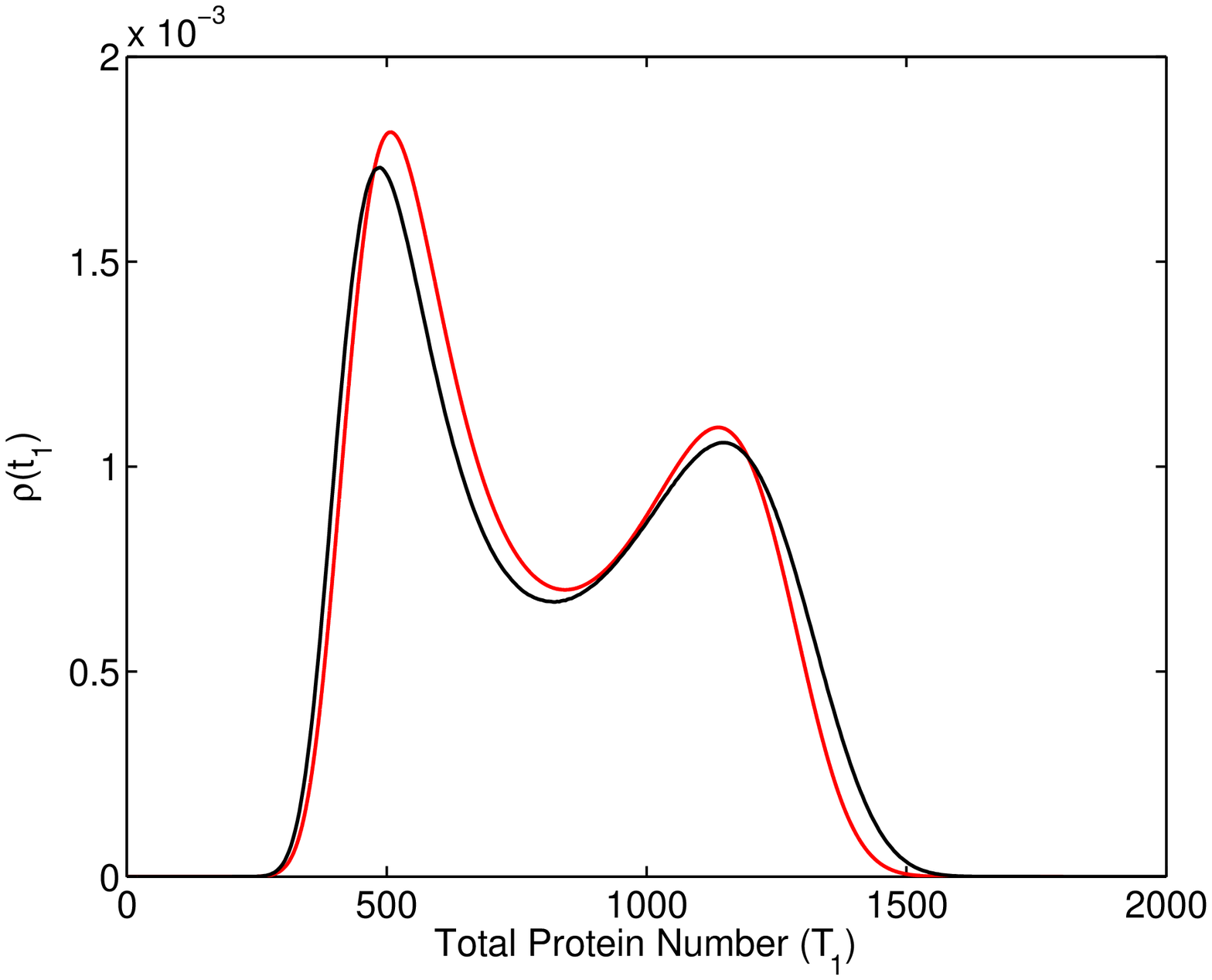}{2in}
\caption{(a) {\it The steady-state distribution for the total 
protein numbers computed
from long-time Monte Carlo simulations of the full model
$(\ref{prodP1})$ -- $(\ref{onoffO2})$.}
(b) {\it The projection of the 2D 
distribution onto the $T_1$ axis (black curve). The red curve is the result of the
equation-free analysis. The parameter values 
used to compute these figures are $\gamma_1 = \gamma_2 = 1.14$,
$\delta_1 = \delta_2 = 7.5 \times 10^4$, $\epsilon_1 = \epsilon_2 = 0,$
$k_1 = k_2 = 5 \times 10^{-4}$, $k_{-1} = k_{-2}= 1$, $k_{o1} = k_{o2} = 0.004$ 
and $k_{-o1}= k_{-02} = 0.1$. These values are consistent with the parameter values
$K = 0.1$, $\delta = 7.5 \times 10^4$
and $\omega = 2 \times 10^{-6}$ used in Stochastic Models I and II.}}
\label{full_model}
\end{figure}

Encouraged by the success of our computational framework
for the simple Stochastic Models I and II considered above, we next
investigated how well these methods would work on the full system 
described by equations (\ref{prodP1}) -- (\ref{onoffO2}). 
We first performed long time 
Monte Carlo simulations using BioNetS \cite{Adalsteinsson:2004:BNS}. 
A two dimensional histogram for the total protein numbers $T_1 =
P_1 + 2 \overline{P_1P_1}$ and $T_2 =
P_2 + 2 \overline{P_2P_2}$
is shown in Figure \ref{full_model}(a). This simulation took 
over 500 total CPU hours 
and is the result of 800 runs distributed over 18 CPUs 
(over a trillion Gillespie SSA steps).  
The black curve in Figure \ref{full_model}(b)
is the projection of the histogram onto the $T_1$ axis. 
We next performed 
equation-free computations for the system. 
As our single observable, $Q$, we used the total 
protein number $T_1$, because this is a quantity that can be measured 
using single cell fluorescent techniques. 
We used a slightly
modified version of step (A3) to compute the conditional density $P({\bf r}|T_1 = t_1)$.
For a given value of $T_1$, we set the rate constants for synthesis and degradation
of this protein equal to zero.
We then ran the simulations for a time of $1 \times 10^5$ to remove any transients. 
Next still keeping $T_1$ fixed, $10000$ samples of the other variables were collected 
at evenly space intervals over a time period of $2 \times 10^5$ and used to generate 
the conditional density. 
A time step of 
$\Delta t = 15$ was used in step (B) of the algorithm. 
To compute the steady state
distribution, polynomials were fit to the average velocity and effective diffusion 
coefficient computed from 
the equation-free analysis and then used to compute the
effective free energy. 
The red curve shown in  Figure \ref{full_model}(b) is the result of the equation-free analysis.
It took less than an hour of CPU time. 
Very good agreement 
between the equation-free method and Monte Carlo simulation is seen. 
Our investigations into these methods revealed that whereas the velocity, $V(q)$,  
is robust to changes in $\Delta t$, the effective diffusion coefficient, $D(q)$,  
is quite sensitive and needs to be treated with care. 
Also, because of
the exponential in the integral for the effective potential, small changes
in the average velocity or effective diffusion coefficient can have large
effects on the steady state distribution. 
Therefore, it is important
to average over sufficiently many realizations to ensure convergence of average
velocity and effective diffusion coefficient. 
Better estimation techniques, such as those developed by A\"{\i}t-Sahalia using
maximum likelihood \cite{AitSahalia:2002:MLE} should be incorporated 
in the data processing step of the algorithms.
Even with these caveats, the 
results presented in this section demonstrate the feasibility and high potential of 
equation-free methods for analyzing stochastic models of genetic networks. 

\subsection{First Passage Time}

\label{secexittimes}

When $\gamma$ is sufficiently large the system
is bistable. 
An important characterization of bistable systems
is the average time for noise-induced transitions between
the stable states. 
Here we make use of the definition of the first
passage time from Section \ref{mathfbt}. 
For the results presented
in this section we use $Q = P_1 - P_2$
as our observable and Stochastic Model I.
The system is bistable for $\gamma > 1.06$.
Let the deterministic stable steady states of $P_1$ be denoted as 
$p_m$ and $p_M$ with $p_m < p_M.$ 
Because of the symmetry of our problem,
$p_m$ and $p_M$ are also the stable
steady states of $P_2.$ 
Let the random variable
${\cal T}_e$ be defined as the first time when ``$P_1 = P_2$" given
the initial conditions $P_1 =p_m$ and $P_2 = p_M.$
In terms of $Q$, this means that ${\cal T}_e$ denotes the time to reach
$Q=0$ when the process starts with $Q$ equal to the negative
steady state $q_m \equiv p_m - p_M$. 
Let $\tau_e$ denote the average of
${\cal T}_e$.  
Then, direct Monte Carlo simulations can be used to
compute the value of $\tau_e.$ The results of such simulations
for three different values of $\gamma$ are presented in the Table 
\ref{tab1}.

\begin{table}
\centerline{
\begin{tabular}{|c|c|c|c|c|c|} \hline
$\gamma$ & $p_m$  & $p_M$ & $q_m = p_m-p_M$ &computed $\tau_e$ from simulations
& $N$  \\
\hline
1.14 & 481.1 & 1038.6 & -557.5 & $7.0 \times 10^5 \pm 6.7 \times 10^3$ & 10 000 \\
\hline
1.20 & 425.8 & 1174.2 & -748.4 & $1.6 \times 10^7 \pm 1.6 \times 10^5$ &  10 000 \\
\hline
1.25 & 392.4 & 1274.3 & -881.9 & $1.0 \times 10^9 \pm 6.3 \times 10^7$ & 250 \\
\hline
\end{tabular}}
\caption{{\it The mean first passage time computed from long-time 
stochastic simulations, averaging over $N$ transitions. The results 
are expressed in the form} 
([sample mean] $\pm$ [sample variance]$/\sqrt{N}$).}
\label{tab1}
\end{table}

\noindent
As expected, the computational time needed to compute the
mean first passage time increases rapidly with $\gamma$. 
In Section \ref{mathfbt}, we introduced two 
formulas (\ref{formulatauep}) and (\ref{tauKramer})
to compute $\tau_e$.
Both formulas  make use of the effective free energy computed by the 
equation-free algorithm.
These potentials
for $\gamma=1.14$, $\gamma=1.20$ and $\gamma=1.25$ are given
in Figure \ref{PhiQsymmetric}.
Consequently, we can compare the results
obtained by the long simulations with the results found from 
formulas (\ref{formulatauep}) and (\ref{tauKramer}) for
$\tau_{e;p}$ and $\tau_{e;k},$ respectively. 
The results
are shown in Table \ref{tab2}.

\begin{table}
\centerline{
\begin{tabular}{|c|c|c|c|} \hline
$\gamma$
& $\tau_e$ from Table \ref{tab1}
& $\tau_{e;p}$  given by (\ref{formulatauep})
& $\tau_{e;k}$  given by (\ref{tauKramer}) \\
\hline
1.14 & $7.0 \times 10^5$ & $6.1 \times 10^5$ & $1.3 \times 10^6$ \\
\hline
1.20 & $1.6 \times 10^7$ & $1.4 \times 10^7$ & $2.6 \times 10^7$ \\
\hline
1.25 & $1.0 \times 10^9$ & $6.7 \times 10^8$ & $1.2 \times 10^9$ \\
\hline
\end{tabular}}
\caption{{\it Comparison of the mean first passage time computed 
from equation-free analysis with long-time stochastic simulations.}}
\label{tab2}
\end{table}

\noindent
Not surprisingly, the results given by $\tau_{e;p}$ are better than
results given by the Kramers' approximation $\tau_{e;k}$. 
However both
methods produce results that are within a factor of 2 of the waiting
times estimated from Monte Carlo simulations. 
As $\gamma$ becomes large
the Monte Carlo simulations become computationally expensive. 
Therefore
only $250$ realizations were used to estimate the mean first passage time,
and we expect that the discrepancy between the Monte Carlo simulations and equation-free
analysis for this case is due to finite sampling errors. 
Initializing the simulation at conditions that are rarely visited by
the direct simulation itself constitutes a form of bias; this bias is
designed to give faster computational estimates of the effective potential
and -- through this -- of the first passage times.
Clearly, this approach hinges on knowledge of a good observable, and in principle
does not depend strongly on the value of the parameter $\gamma$;
therefore, the larger the parameter $\gamma$ the higher the computational speedup
in the first passage time estimation that will result.
A quantitative study of this speedup is underway and will be reported
elsewhere; it does not lie within the scope of this paper.
We stress, however, that
(as in molecular dynamics simulations) knowledge of a good observable (a good
``reaction coordinate") is crucial for the success of the approach.

Note that formula $\tau_{e;k}$ requires estimates of
the second derivative
of the potential at points $q_u$ and $q_m.$ 
To do this, we fit $\Phi(q)$ locally to
a polynomial and used
the derivatives of the polynomial at the required points;
once again, maximum likelihood techniques (e.g. \cite{AitSahalia:2002:MLE})
should be used for better results. 
The formula for $\tau_{e;p}$,
requires the evaluation of an
indefinite integral. 
The integral was approximated by considering
only a finite interval that neglected contributions
from the region  of sufficiently small $q$
where the potential $\Phi$ is very large.

\subsection{Bifurcations}

\label{secbifur}

\begin{figure}
 \centerline{
 \psfig{file=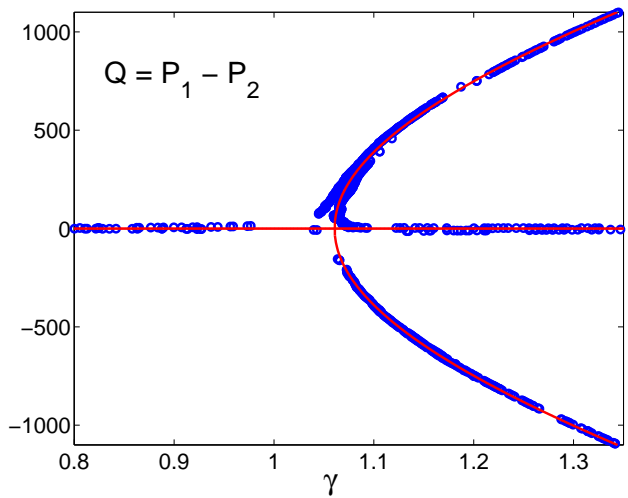,height=2.6in}
 }
 \caption{{\it A plot of the steady states obtained by equation-free 
 analysis $(\ref{numcont})$ (blue circles). Also shown
 are the deterministic steady states from Figure $\ref{figQgammadeter}$
 (red line).}}
 \label{figurebifur}
 \end{figure}

In this section, our goal is to
run the simulations for short times only and
compute a form of ``stochastic bifurcation diagram" 
using continuation methods, as an extension of the deterministic 
bifurcation computations. 
We use Stochastic Model I and study
the dependence of the ``steady states" on $\gamma$; the ``steady
states" we report are the fixed points of the algorithm from 
Section \ref{secbifth} with
the conditional density $P(R|Q=q)$ approximated
by the Dirac delta function in ($\mathbb A$) -- ($\mathbb B$),
similar to the approach (A1) from Section
\ref{secmathfram}. 
Numerical results are given in Figure \ref{figurebifur}. 
For comparison we also plot 
the steady states of the corresponding deterministic equation (compare with
Figure \ref{figQgammadeter}). 
The plot in Figure \ref{figurebifur} was computed by initializing 
on different branches far from the bifurcation point and continuing 
from these different initializations (our simple arclength continuation
algorithm did not include a ``pitchfork detection" component).

The accuracy of the numerical results depend on several factors: 
the estimation technique for the Jacobian elements,
the tolerance of the error for Newton-Raphson iterations, 
the number of realizations which are used to evaluate $F$, 
the time interval $\Delta t$ and the steepness of the underlying potential $\Phi$.
As can be seen in Figure \ref{figurebifur}, stochasticity along with 
all these numerical factors
have slightly perturbed the pitchfork bifurcation; this could be exacerbated
by our choice of (symmetric or asymmetric) observable.
It is easy to follow any branch of steady states far from the bifurcation point. 
For obvious reasons this becomes more complicated when we
are close to the ``bifurcation point" at $\gamma=1.06$. 
The main problem is that the potential becomes
``flat" close to the bifurcation point --- see Figure \ref{PhiQsymmetric}. 
One way to improve the results is to adaptively change the number of realizations
in ($\mathbb B$). 
That is, if the Newton-Raphson iterations of (\ref{numcont}) do not
converge to a desired tolerance, then more realizations are
added. 
Another approach is to estimate directly a local polynomial model of
the underlying diffusion process from discrete SSA data 
using maximum likelihood
tools, and then search for the bifurcations of the critical points
of the effective potential.
Indeed, one can plot the zeroes of the estimated drift, or -- in the case
of a state-dependent diffusion coefficient -- one can correct them to report
the maxima of the steady state distribution 
\cite{Kopelevich:2005:CGK,Sriraman:2005:CND}; 
both of these are good
candidate bifurcation diagrams for the stochastic case.
When the potential is steep and the equilibrium is ``less noisy" it is not necessary 
to use many realizations; the relation between computational effort (in terms of
number of replicas, simulation time horizon and estimation method) and resulting
accuracy is, again, a subject of current investigation beyond the scope of
this paper.

\begin{figure}
 \centerline{
 \psfig{file=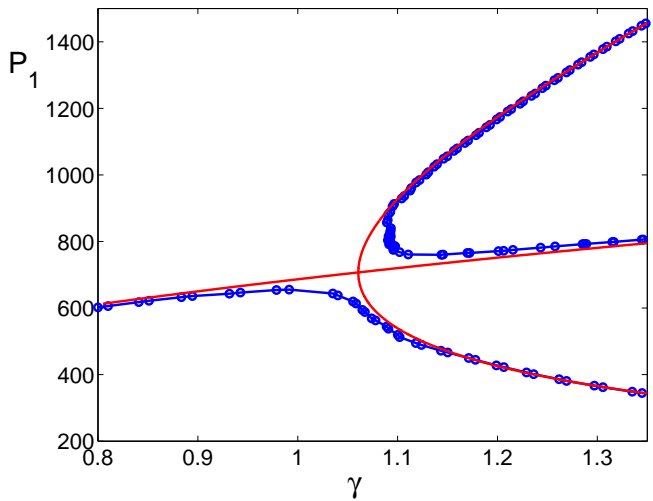,height=2.6in}
 }
 \caption{{\it Plot of the steady states obtained by 
 using $P_1$ as the observable (blue circles). Again for comparison the 
 deterministic case is shown as the red line.}}
 \label{figbifurP1}
 \end{figure}

Finally, the results using  
$P_1$ instead of $Q$ as the observable are shown in Figure \ref{figbifurP1}. 
In this case, the asymmetry of our observable, and the perturbation 
it causes on the initialization process, make the perturbation of
the pitchfork bifurcation stronger.
Of course, the results depend on the initialization procedure, 
our estimation technique, the error tolerance, the number of 
realizations, the length of time step $\Delta t$ as well as the type
of continuation algorithm we are using (here we used a very simple
one, without bifurcation detection, in order to demonstrate what is
possible). 
Accurate bifurcation detection depends on accurate Jacobians and even
higher derivatives; estimating these from dynamic (and noisy !) data 
is notoriously difficult. 
While conceptually we do have the tools to ``hone in" the more accurate 
detection of bifurcation points, careful quantitative work is necessary 
to pin down the tradeoffs between computational effort, model estimation accuracy
and bifurcation point estimation accuracy.

\section{Discussion}

\label{secdiscussion}

In this paper we discussed and illustrated the use of
certain equation-free numerical techniques that have the potential
to accelerate the computer-assisted analysis of stochastic models
of regulatory networks.
There is a clear current need for accelerating such simulations:
even for modestly complex regulatory networks, stochastic models rapidly
become computationally expensive.  
Computational acceleration is usually based on model reduction; 
theoretical methods for stochastic model reduction that take advantage of a
separation of time scales are the focus of intense current research 
\cite{Kepler:2001:STR,Cao:2005:SCS,Haseltine:2002:ASC,Rao:2003:SCK,Salis:2005:AHS}. 
As we discussed in the introduction, many important gene regulatory
networks do satisfy this assumption of a separation of time scales
because synthesis and degradation of new proteins and transcripts usually
occurs on a slower time scale than processes that change the chemical
state of proteins.
Analytical model reduction techniques  assume
that the fast variables are in quasi-steady state with respect to the
slow variables, and use the quasi-steady state distributions conditioned on the
slow variables to eliminate the fast variables by averaging.
These methods have been successfully applied to simple models, but the
theory is not as well established as the deterministic
counterpart. 
Having an explicit model lies often at the basis of such stochastic reduction 
methods.

In the equation-free approach many of the same elements (separation
of time scales, approximation of conditional quasi-steady distributions)
also underpin computational efficiency; but the basic premise is that 
the model is available in the form of a ``black box" simulation code.
We do not try to first reduce, and then simulate the reduced surrogate;
we try to design the smallest number of ``intelligent" short 
computational experiments with the full stochastic model to find the
quantities of interest, whether these are steady state probability
distributions, their maxima, or transition rates in the bistable case.
In that sense, the approaches we described here {\it do not} hinge 
upon the ``inner", detailed simulator as being a Gillespie SSA one - 
the methods are equally applicable to any ``inner" simulator, stochastic
or deterministic, as long as the main assumption of a low-dimensional
{\it effective} stochastic model is a good one for the long-term
system dynamics.
Indeed, if another reduction method can be used to produce a good
approximate dynamic simulator, our algorithms can be ``wrapped" around
this surrogate simulator rather than the full model for further acceleration.

Another important point has to do with the type of computation we
are interested in - do we want to accelerate the {\it direct simulation}
of the model, or do we want to accelerate the computation of certain
features of its long-term dynamics (e.g. of the maxima of the steady state
distribution)?
These latter quantities can be also obtained from long-term direct simulation,
but one of the points that we want to stress is that we can link direct
simulation to {\it different} numerical algorithms (such as contraction mappings,
and continuation methods) to obtain these quantities, often faster than with
direct simulation alone.
In the same way that bifurcation diagrams for dynamical systems are usually 
{\it not computed} through direct ODE integration, but through bifurcation
algorithms, the parametric dependence of the long-term dynamics of stochastic
models does not have to be computed through long-time direct simulation only.
This ``alternative" acceleration, not through accelerating the direct simulation
itself, but through linking it to different numerical algorithms, lies at the
basis of the equation-free framework.

Having said this, we briefly mention that equation-free methods for accelerating
the direct simulation itself also exist.
{\it Coarse projective integration}
which uses short bursts of direct simulation to estimate time derivatives of
evolving probability densities, and then passes them to standard numerical
integration algorithms, has been successfully used in many contexts 
\cite{Kevrekidis:2003:EFM,Erban:2004:CAE,Gear:2001:PIM,Gear:2002:CIB}
Coarse projective integration has a strong relation to the direct
simulation acceleration methods in \cite{Gillespie:2001:AAS}; it has not been 
discussed 
in this paper, because we chose to focus on {\it very long-term} features of
the network dynamics; it might interest the reader that the method can be
used to also integrate backward in time, and solve ``effective boundary value
problems" to find  ``coarse" limit cycles \cite{Martinez:2004:CPK}.

In the equation-free methods for analyzing stochastic models
of gene regulation that we discussed in this paper, we have
tried to circumvent the difficulties encountered by direct
simulation (in this case SSA) through the design of short bursts
of {\it appropriately initialized} computational experiments
with the full simulator.
In a sense, we ``resign ourselves" to the fact that the direct
simulator is expensive; we ask what is the shortest amount of
running of this expensive direct simulator in order to obtain the
quantities we are interested in.
The ``design of experiment" protocols are  templated on traditional
continuum numerical methods, like the fixed point and continuation
algorithms to compute bifurcation points, or quadrature to estimate
Kramers' formula.
The only difference is that the quantities (residuals, actions of
Jacobians, values of the integrand) that are required for numerical
computation are not given by a closed formula, but rather through
direct numerical simulation of the full model and estimation.
We reiterate once more that these techniques can be wrapped around the
full direct simulator, or our best available reduction of it, 
without change.

Knowing appropriate coarse-grained observables (the variables
in terms of which the unavailable effective model would be written)
is an important feature of the algorithms.
Extensive experience with the problem, intuition, or analytical
work may often suggest such observables; we did take advantage of
such knowledge in this paper.
We did already demonstrate an important point: more than one observables
are capable of doing a good job as the parameterizing variables in an
equation-free context; one does not need to know {\it the exact} slow
variables.
This issue is discussed extensively for the deterministic context in
\cite{Gear:2004:PSM}.
It is, however, important to note that algorithms for the detection of
low-dimensionality in high-dimensional data can be vital in {\it suggesting}
such observables from simulations.
Principal Component Analysis is an established linear method for the detection
of appropriate lower-order observables from simulation data; numerically
estimated eigenmodes of the problem may also provide good observables
(see the discussion in \cite{Siettos:2003:CBD} about estimating gaps between
eigenvalues, and using them to decide whether we should include more 
observables as independent variables).
There are, however, some important developments in this area: the recent 
use of harmonic analysis (geometric diffusion) on graphs constructed from
high-dimensional data shows great promise in detecting good observables
(reaction coordinates) for complex, high dimensional systems 
\cite{Nadler:2005:DMS,Belkin:2003:EDR,Nadler:2005:DMS2}.
This ``variable-free" approach can be naturally linked to equation-free
computation (one designs computational experiments both to detect the
appropriate observables {\it and} to do computations with them) 
\cite{Nadler:2005:DMS};
we are currently working on demonstrating this link for gene regulatory 
network modeling. 

It is clear that, in certain cases, an equation-free computational
approach is expected to have advantages over direct simulation.
For steep potentials and low noise, for example, the way equation-free
computation uses a good observable to bias the simulation will sample
the effective potential and give a good estimate of the transition rates
much faster than direct simulation.
Also, parametric analysis methods should be able to explore parametric
transitions faster and more systematically than direct simulation, in 
analogy with the use of bifurcation techniques rather than direct
simulation in deterministic dynamical systems (e.g. by writing augmented 
algorithms that converge on marginally stable or unstable solutions).
The complexity of the computation depends crucially on the dimensionality
of the unavailable {\it reduced} model, and not so crucially on the 
dimensionality of the detailed, full model.
We are currently working on the quantification of these computational
benefits; this work is complicated by the fact that -- lacking explicit
formulas from which to obtain derivative information -- errors must
be computed on-line, through {\it a posteriori} estimates.

This brings us to a final, yet vital issue: estimation. 
Given the noisy nature of the data, estimating the numerical quantities
of interest lies at the heart of the accuracy (and thus the viability)
of the computation.
For our gene networks, these quantities included
the effective potential $\Phi(q)$ and the effective
diffusion coefficient $D(q)$. 
Preliminary investigations
revealed that the effective diffusion coefficient, $D(q)$,
is quite sensitive to the time step $\Delta t$ and needs to be treated
with care.
Also, small changes in the average velocity or effective diffusion
coefficient
can have relatively large effects on the steady state distribution.
Even though some computations are ``embarrassingly parallel" (one 
short, fine scale realization per processor, running independently)
variance reduction becomes an important feature (see, e.g. 
\cite{Melchior:1995:VRS}).
Maximum likelihood estimation techniques (e.g. \cite{AitSahalia:2002:MLE}) 
take the place of simple formulas such as  
(\ref{avgvel}) and (\ref{effdiff});
one can envision certain hypothesis testing computations
(is our model locally well-approximated by a diffusion process ?)
becoming part of the overall computational scheme.
Until these elements, and their computational cost, are analyzed
and tested, there will be no firm guarantees for the computational 
efficiency of equation-free methods.
Yet, even with these caveats,
as we computationally demonstrated in this paper, we believe that 
the equation-free framework provides a promising new approach
to gene regulatory network modeling, alternative
to long-direct simulation.
It links directly with powerful and tested traditional continuum 
numerical algorithms (such as numerical integration, fixed point 
algorithms, matrix-free iterative linear algebra) and with 
system theory techniques like filtering and estimation.
These
techniques are, in some sense ``off the shelf" and do not need to
be redeveloped.
In our opinion, it is the linking of equation-free techniques with
novel data reduction/clustering techniques (such as the use of the
graph Laplacian to detect good reaction coordinates \cite{Nadler:2005:DMS})
that hold the most promise in the computational study of complicated
stochastic systems in general, and of gene regulatory networks and their
models in particular.

\noindent
{\bf Acknowledgments.} This work was partially supported by DARPA and 
an NSF/ITR grant (I.G.K). This work was partially supported by Biotechnology 
and Biological Sciences Research Council (R.E.). This work was supported 
by DARPA (F30602-01-2-0579) (T.C.E.).

\bibliographystyle{amsplain}
\bibliography{bibrad}

\end{document}